\documentclass[structabstract]{aa}  
\newcommand{\mkref}[1]{$\mathrm{^{#1}}$}

\usepackage{color}
\usepackage{graphicx}
\usepackage{natbib}
\bibpunct{(}{)}{;}{a}{}{,} %
\usepackage{txfonts}
\begin{document}
   \title{On the alignment between the circumstellar disks and orbital planes of Herbig Ae/Be binary systems}

   \author{H.E. Wheelwright
          \inst{1}
          \and
          J.S. Vink\inst{2}
	  \and	
	  R.D. Oudmaijer\inst{1}
          \and 
          J.E. Drew\inst{3}}
   
   \institute{School of Physics and Astronomy, University of Leeds, Leeds
     LS2 9JT, UK\\
     \email{H.E.Wheelwright@leeds.ac.uk}
     \and  Armagh Observatory, College Hill, Armagh BT61 9DG, Northern
     Ireland
     \and Centre for Astrophysics Research, STRI, University of Hertfordshire, College Lane Campus, Hatfield AL10 9AB, UK
}

   \date{Received March 31, 2011; accepted June 09, 2011}

 
  \abstract
{The majority of the intermediate mass, pre-main-sequence Herbig Ae/Be stars
  reside in binary systems. As these systems are young, their properties may
  contain an imprint of the star formation process at intermediate masses
  (2-15\,${M_{\odot}}$). However, these systems are generally spatially
  unresolved, making it difficult to probe their circumstellar environment to
  search for manifestations of their formation process, such as accretion
  disks.}
{Here we investigate the formation mechanism of Herbig Ae/Be (HAe/Be) binary
  systems by studying the relative orientation of their binary orbits and circumstellar disks.}
{We present linear spectropolarimetric observations of HAe/Be stars
  over the H$\alpha$ line, which are used to determine the orientation
  of their circumstellar disks. In conjunction with data from the
  literature, we obtain a sample of 20 binaries with known disk
  position angles (PAs). We subsequently compare our disk PA data to a
  model to investigate whether HAe/Be binary systems and their disks
  are co-planar. Moreover, in the light of a relatively recent
  suggestion that some HAe/Be star spectropolarimetric signatures may
  not necessarily be related to circumstellar disks, we re-assess the
  relationship between spectropolarimetric signatures and disk PAs. We
  do this {{\color{black}{{by}}}} comparing spectropolarimetric and high spatial resolution
  observations of young stellar objects (both HAe/Be and T Tauri
  stars).}
{We find that spectropolarimetric observations of pre-main-sequence
  stars do indeed trace circumstellar disks. This finding is
  significant above the 3$\sigma$ level. In addition, our data are
  entirely consistent with the situation in which HAe/Be binary
  systems and circumstellar disks are co-planar, while random
  orientations can be rejected at the 2.2$\sigma$ level.}
 {The conclusive alignment (at more than 3$\sigma$) between the disk
   PAs derived from linear spectropolarimetry and high spatial
   resolution observations indicates that linear spectropolarimetry
   traces disks. This in turn allows us to conclude that the orbital
   planes of HAe/Be binary systems and the disks around the primaries
   are likely to be co-planar, which is consistent with the notion
   that these systems form via monolithic collapse and subsequent disk
   fragmentation.}

\keywords{Techniques: polarimetric -- Stars: emission-line
   --  formation -- binaries -- pre-main-sequence -- variables: T
   Tauri, Herbig Ae/Be} 
\titlerunning{The relative alignment of Herbig Ae/Be binaries and disks}
\authorrunning{H.E. Wheelwright et   al.} 
 \maketitle
%

\section{Introduction}

Herbig Ae/Be stars are intermediate mass, $\mathrm{2-15}\,M_{\odot}$,
pre-main-sequence stars \citep[see
e.g.][]{Herbig1960,Hillenbrand1992,Waters1998,Hernandez2004}. They are
the most massive objects to experience an optically visible
pre-main-sequence (PMS) phase. Therefore, they provide a key
opportunity to study the early evolution of stars more massive than
the lower mass T Tauri stars at these wavelengths. Herbig Ae/Be
(HAe/Be) stars span the transition between low and high mass
objects. As a result, it has been postulated that there is a change in
accretion mechanism within the HAe/Be mass range \citep[see
e.g.][]{Vink2002,Eisner2004,Monnier2005}. If this is the case, the
more massive Herbig Ae/Be stars may form in the same fashion as the
most massive stars. Consequently, the study of Herbig Ae/Be stars may
be used to constrain the process of massive star formation, which is
still not fully understood \citep[see e.g.][]{ZinneckerandYorke2007}.

\smallskip

\citet{DB2006} used spectroastrometry to detect unresolved HAe/Be
binary systems and they compared the position angle (PA) of the
systems to the orientation of their circumprimary disks. They
reported that the binary and circumprimary disk PAs of the systems in
their sample are preferentially aligned. This indicates that the
circumstellar disks lie in the same plane as the binary orbit. In
turn, this suggests that these systems formed via a scenario that
features fragmentation, as opposed to alternatives involving the
capture of binary companions.

\smallskip

This is consistent with one model of massive star formation. Using 3D
radiation-hydrodynamic simulations, \citet[][]{Krumholz2009} show that
stars of even $\mathrm{40}M_{\odot}$ can form via monolithic collapse
and disk accretion. A prediction of these models is that disk
fragmentation leads to binary systems. The resultant binary system is
aligned with the original fragmented disk structure. Therefore,
because of angular momentum considerations, the binary system is also
aligned with the inner accretion disk that reaches onto the primary
star's surface. In addition, the secondary, at a distance of
approximately 1000~AU, has a relatively high mass. It has also been
proposed that massive stars form via competitive accretion. In this
scenario, a star gains material from its parental cluster. In the
process, binary systems are formed via stellar capture. These binaries
start as solar mass binaries at separations of the order of 1000~AU
and evolve into close high mass binaries at approximately 1~AU
\citep[][]{Bonnell2005}, much closer than in the monolithic accretion
scenario. Also, as these systems are formed via stellar capture, the
binary orbital planes and circumprimary disks do not necessarily lie
in the same plane \citep[]{JBally2007}, unlike the monolithic collapse
and disk fragmentation scenario. Therefore, the properties of binary
systems can be used to constrain the star formation mechanism at high
masses.

\smallskip

The work of \citet{DB2006} was extended by \citet{Wheelwright2010} who
used spectroastrometry to determine the mass ratios of HAe/Be binary
systems. They found that the mass-ratio distribution of HAe/Be binary
systems favours comparable masses, further indicating that these
systems formed via disk fragmentation. However, the sample of
\citet{DB2006} which initially suggested this was small (6 objects),
while a comparison with model simulations of co-planar binary orbits
and disks was lacking. Therefore, the relative alignment of HAe/Be
binary orbits and circumstellar disks is still uncertain. We aim to
extend the work of \citet{DB2006} and address this uncertainty by
enlarging the sample of binary HAe/Be stars with known disk PAs.

\smallskip

To achieve this, we require a large sample of HAe/Be binary systems
with known disk orientations. We generate a sample of 20 such systems
using linear spectropolarimetry and high spatial resolution data from
the literature. Spectropolarimetry offers a unique opportunity to
probe the circumstellar environment of young stellar objects on scales
small enough to study accretion disks. If a star is surrounded by a
hot disk, free electrons in the disk can polarise the light of the
central star. Emission-line photons emanating from the disk do not
pass through the polarising medium, unlike the continuum
light. Consequently, emission lines are depolarised with respect to
the continuum. Therefore, a depolarisation signature over an emission
line can be used to infer the presence of a small-scale, otherwise
undetected disk
\citep{Clarke1974,Poeckert1975,Poeckert1976}. Alternatively, if
accretion shocks produce emission lines close to the star, the
presence of a disk may be manifest by scattering polarisation of a
compact source of line emission \citep[][hereafter
V2002]{Vink2002}. Furthermore, spectropolarimetric signatures can also
be used to constrain the geometry of such disks
\citep[e.g.][]{DisksVink2005}.

\smallskip

Linear spectropolarimetry is well established as a technique to study
spatially unresolved disks around HAe/Be stars \citep[see
e.g.][]{Oudmaijer1999,Vink2002,Vink2005a,JCMottram2007}. However, the
use of spectropolarimetry to study disks has recently been revisited
by \citet[]{Harrington2007}. The authors report observations of
spectropolarimetric signatures across absorption components of
H$\alpha$ emission lines. It is suggested such signatures are due to a
process other than scattering in a disk. Instead, \citet{Kuhn2007}
propose that optical pumping and absorption in winds may be
responsible. Optical pumping, i.e. a preferential population of
different magnetic sub-states of the lower level of an atomic
transition, may be caused by an anisotropic radiation field.  If a gas
is optically pumped, its opacity will depend upon the orientation of
the electric field of incident radiation. As a result, the absorption
of light in such gas can result in linear polarisation effects across
absorption features \citep[see][]{Kuhn2007,Kuhn2010}. We note that all
spectropolarimetric signatures require a flattened, asymmetric
geometry otherwise the polarisation vectors cancel. This is
independent of the polarising mechanism. Nonetheless, if the
polarisation signatures of HAe/Be stars originate in significantly
asymmetric winds, this does not automatically demand an inward
coplanar extension of any large-scale circumstellar disk. Here we
investigate whether there is a clear relationship between the angle of
polarisation and disk orientation which, at minimum signals alignment
between the unresolved and resolved spatial scales. Such an outcome is
required if indeed the bulk of the observed linear polarisation of
these objects is due to a circumstellar disk.

\smallskip

This paper is structured as follows. In Section \ref{disk_bin} we
evaluate the alignment of HAe/Be binary systems and circumstellar
disks using linear spectropolarimetry over H$\alpha$ to determine the
PAs of disks in HAe/Be binary systems. To justify the use
of spectropolarimetry to trace circumstellar disks, we reassess the
relationship between spectropolarimetric signatures and disk PAs in
Section~\ref{initial_test}. Finally, we discuss the results and
conclude the paper in Section~\ref{disc}.

\section{The relative orientation of HAe/Be binary systems and circumstellar disks.}

\label{disk_bin}
The key to determining whether HAe/Be binary planes and circumstellar
disks are preferentially aligned is to use a large sample of binary
systems with known disk PAs. There are several studies of HAe/Be star
binary systems \citep[see e.g.][]{DB2006,Wheelwright2010}. However,
there are few measurements of the orientation of the circumstellar
disks in these systems. Linear spectropolarimetry is the favoured
technique to probe the orientation of HAe/Be star disks since a sample
of $\sim$20 can be obtained in a few nights of observing (see
e.g. V2002). To maximise our sample of disk PAs, we have conducted new
H$\alpha$ spectropolarimetric observations; taken linear
spectropolarimetric results from previous work; and assimilated
additional constraints on HAe/Be star disk orientations from the
literature. We first present the additional spectropolarimetric
observations we undertook (Section~\ref{obs}). We then use the
assembled sample of disk PAs, in conjunction with a simple model
(described in Section \ref{model}), to evaluate the hypothesis that
circumstellar disks in HAe/Be binary systems lie in the orbital plane
(Section \ref{alignment}).

\subsection{Linear spectropolarimetric observations}
\label{obs}

We present linear spectropolarimetric observations of HAe/Be stars
conducted in the \textit{R} band and centred on
H${\alpha}$. Here we describe the observing procedure and the
data reduction steps before presenting the observed
spectropolarimetric signatures.

\smallskip

The sample was predominately chosen from the catalogue of
\citet{The1994}. Targets were selected based on their visual magnitude
(\textit{V} $\rm \leq $ 12-13). This is because spectropolarimetry
requires high signal-to-noise-ratio (SNR) data and thus only bright
targets can be observed. Most of the targets were binary Herbig Ae/Be
stars that had not been observed with spectropolarimetry previously,
and these were primarily drawn from \citet{DB2006} and
\citet{Thomas2007}. Several objects with existing spectropolarimetric
data were observed to check the results are consistent with previous
observations.

The linear spectropolarimetric data were obtained using the William
Herschel Telescope (WHT) from 08-11-2008 to 10-11-2008. Clouds were
present for the majority of the three nights, preventing observation
for some, but not all, of the time. The seeing was typically fair
($\sim$1\arcsec), although on occasions it became poor
(2--2.5\arcsec). The observations were conducted with the ISIS
spectrograph which was equipped with polarising optics comprising of a
calcite block and a rotating half-wave plate. The R1200R grating was
used and the central wavelength was set to
6560~${\AA}$. Several slit widths were used, ranging from 1 to
1.8 arcsec, and the minimum spectral resolution was found to be
$\mathrm\sim4250$, or 70~$\mathrm{km\,s^{-1}}$.

\smallskip

The calcite block separated the incident light into two perpendicular
rays: the ordinary (O) and the extraordinary (E) rays. Each
observation comprised of both the O and the E ray spectrum of the
science target, and a corresponding set of sky spectra. The
polarisation at PAs of
$\mathrm{0{\degr},\,22.5{\degr},\,45{\degr}\, and\, 67.5{\degr}}$ was
measured by rotating the half-wave plate. Multiple polarisation
standard stars were observed to characterise the instrumental
polarisation and calibrate the polarisation angle. A log of the
observations is presented in Table \ref{spec_pol_obs}.

\begin{center}
  \begin{table*}
    \begin{center}
      \caption{The log of observations. The continuum polarisation was measured in the wavelength region 6520--6600~$\AA$, excluding changes over the H$\alpha$ line. The uncertainty in the continuum polarisation is typically 0.1 per cent and the uncertainty in the continuum polarisation angle is of the order of $\mathrm{1{\degr}}$. The polarisation angles are in the equatorial frame.\label{spec_pol_obs}}
      \begin{tabular}{l l r r l r r l l l r r}

      \hline 
      Object & Alt. Name & \multicolumn{1}{c}{RA} & \multicolumn{1}{c}{Dec} & Spec. Type & \multicolumn{1}{c}{$V$} & \multicolumn{1}{c}{$\mathrm{T_{exp}}$} & Slit & Seeing & Date & \multicolumn{1}{c}{$\mathrm{P_{cont}}$} & \multicolumn{1}{c}{$\mathrm{\theta_{cont}}$}\\
 &      & \multicolumn{1}{c}{(J2000)}   & \multicolumn{1}{c}{(J2000)} &   & \multicolumn{1}{c}{(mags)} & \multicolumn{1}{c}{(mins)}& (\arcsec) & (\arcsec)  & & \multicolumn{1}{c}{(\%)} & \multicolumn{1}{c}{($\mathrm{{\degr}}$)}\\
      \hline
      \hline
      \object{XY Per}        &  HD 275877                     & 03 49 36.3       & +38 58 55.5   & A2IIv      & 9.4  & 120.0 & 1.2 & 0.7 & 10-11-2008  &  $\mathrm{1.4}$ & $\mathrm{128}$ \\
      \object{MWC 758}       &  HD 36112                      & 05 30 27.5       & +25 19 57.1   & A5IVe      & 8.3  & 60.0  & 1.5 & 1.0 & 08-11-2008  & $\mathrm{0.5}$ & $\mathrm{47}$ \\
      \object{HK Ori}        &  MWC 497                       & 05 31 28.1       & +12 09 10.2   & A4pev      & 11.7 & 133.3 & 1.5 & 1.5 & 08-11-2008 & $\mathrm{1.4}$ & $\mathrm{113}$\\
      \object{V586 Ori}      &  HD 37258                      & 05 36 59.2       & $-$06 09 16.4 & A2V        & 9.8  & 146.7 & 1.5 & 0.9 & 09-11-2008 & $\mathrm{0.9}$ & $\mathrm{84}$ \\
      \object{BF Ori}        &  HBC 169                       & 05 37 13.3       & $-$06 35 0.6  & A5II-IIIev & 10.3 & 133.3 & 1.5 & 0.8 & 09-11-2008  & $\mathrm{0.6}$ & $\mathrm{58}$  \\
      \object{V350 Ori}      &  HBC 493                       & 05 40 11.8       & $-$09 42 11.1 & A1         & 11.5 & 120.0 & 1.5 & 0.6 & 10-11-2008 & $\mathrm{1.1}$ & $\mathrm{65}$ \\
      \object{MWC 147}       &  HD 259431                     & 06 33 5.2        & +10 19 20.0   & B6pe       & 8.8  & 53.4  & 1.5 & 0.9 & 09-11-2008  & $\mathrm{1.0}$ & $\mathrm{100}$ \\
      \object{GU CMa}        &  HD 52721                      & 07 01 49.5       & $-$11 18 3.3  & B2Vne      & 6.6  & 15.3  & 1.5 & 1.3 & 08-11-2008 & $\mathrm{0.8}$ & $\mathrm{24}$ \\
      \object{HD 179218}     &  MWC 614                       & 19 11 11.3       & +15 47 15.6   & A0IVe      & 7.2  & 60.0  & 1.5 & 1.2 & 08-11-2008 & $\mathrm{0.5}$ & $\mathrm{111}$ \\
      \object{HBC 310}       &  AS 477                        & 21 52 34.1       & +47 13 43.6      & B9.5Ve     & 10.2 & 110.0 & 1.5 & 1.0 & 09-11-2008 & $\mathrm{1.0}$ & $\mathrm{50}$  \\  
      \object{Il Cep}        &  HD 216629                     & 22 53 15.6 & +62 08 45.0      & B2IV-Vne   & 9.3  & 86.7  & 1.5 & 1.3 & 08-11-2008 & $\mathrm{4.3}$ & $\mathrm{100}$ \\
      \object{MWC 1080}      &  V628 Cas                      & 23 17 25.6 & +60 50 43.6      & B0eq       & 11.6 & 96.7  & 1.5 & 1.0 & 09-11-2008  & $\mathrm{1.7}$ & $\mathrm{78}$  \\
      
      \hline
      \end{tabular}

    \end{center}    

  \end{table*}
\end{center}

Data reduction was conducted using the Image Reduction and Analysis
Facility ({\sc{iraf}})\footnote{http://iraf.noao.edu/, see
\citet{IRAF}}, in conjunction with routines written in Interactive
Data Language ({\sc{idl}}). The data reduction process for each
observation consisted of trimming, bias subtraction, flat-field
division and cosmic-ray removal. Following the above, the target O and
E spectra, and those of the sky if they were present, were extracted
from each frame. {\color{black}{{Sky spectra, which were not always
detected, were typically a few percent of the stellar spectra. We note
that polarisation signatures over H$\alpha$, which are the focus of
the paper, are unaffected by contaminant sky polarisation. Therefore,
the sky polarisation has no influence on our final results.}}}
Wavelength calibration was performed using CuNe and CuAr arc spectra,
which were obtained periodically during the observing run.

\smallskip

Once the O and E ray spectra had been extracted, the Stokes parameters
for each data set were calculated using a routine written in
{\sc{idl}}. The method used is that outlined in the ISIS polarisation
manual by Jaap Tinbergen and Ren\'{e}
Rutten\footnote{http://www.ing.iac.es/Astronomy/observing/manuals\\\hspace*{5mm}/html\_manuals/wht\_instr/isispol/isispol.html}. For
each set of polarisation data, i.e. data obtained with the half-wave
plate at $\mathrm{0{\degr}\,and\,45{\degr}}$ or
$\mathrm{22.5{\degr}\,and\,67.5{\degr}}$, the ratio of the O and E
rays in each spectrum was calculated. To obtain the degree of
polarisation, the data obtained at a given PA were averaged and the
following equations were used:

\begin{equation}
R^2=\frac{I_{\mathrm{O,0{\degr}}}/I_{\mathrm{E,0{\degr}}}}{I_{\mathrm{O,45{\degr}}}/I_{\mathrm{E,45{\degr}}}}
\end{equation}

\begin{equation}
q=\frac{R-1}{R+1}
\end{equation}

where $q$ = $Q/I$, $I$ is the total flux input and
${I_{\mathrm{O,ang}}}$ and ${I_{\mathrm{E,ang}}}$ are the fluxes of
the O and E rays at the stated half-wave plate PAs. This procedure was
repeated for the other set of polarisation data, i.e. data obtained
with half-wave-plate PAs of
$\mathrm{22.5{\degr}\,and\,67.5{\degr}}$, to calculate $u$.

\smallskip

To calculate the total polarisation and the polarisation angle the
data were combined using the following equations:
\begin{equation}
P = \sqrt{{{q}^2}+{{u}^2}} 
\end{equation}

\begin{equation}
\label{theta}
\theta = \frac{1}{2}\mathrm{tan^{\mathrm{-1}}}\left(\frac{{u}}{q}\right) 
\end{equation}

 where $P$ represents the total polarisation, and $\theta$ is the
 polarisation angle.

\smallskip

{\color{black}{{Instrumental polarisation was not corrected for. The
standard observations indicate that the instrumental polarisation is
of the order 0.1 per cent, and is not the dominant source of continuum
polarisation. Interstellar polarisation is not corrected for either,
as such corrections are typically subject to significant uncertainties
\citep[see e.g.][]{Jensen2004}. Contaminant polarisation simply adds a
wavelength independent vector to the Stokes $Q$ and $U$ parameters.}}}
Therefore, plotting the spectrally dispersed $q$ against $u$ allows
the intrinsic angle of polarisation, and hence the polarising media,
to be established.

\subsubsection{Spectropolarimetric signatures}

\label{sigs}

We present an example of the spectropolarimetric signatures observed
in Figure \ref{spec_pol_ex}. Data around H$\alpha$ for all the stars
in the sample are presented in Appendix \ref{spec_pol_app} in Figure
\ref{spec_pol_tri}, while continuum polarisation measurements are
included in Table~\ref{spec_pol_obs}. The signatures of the objects
previously observed with spectropolarimetry are generally consistent
with published results
\citep[e.g.][]{Vink2002,Vink2005a,JCMottram2007}. This provides an
important check on the data reduction process. In general, the data
are of slightly inferior quality to previous observations. This is
attributed to the poor weather conditions throughout the
observing. Consequently a coarser binning is used than is typical for
such data. Five objects exhibit a change in both linear polarisation
and the polarisation angle over H$\alpha$ (HD 179218, HK Ori, MWC
1080, V586 Ori \& MWC 147). MWC 1080 and MWC 147 were also observed by
\citet[][hereafter M2007]{JCMottram2007}. While the data are broadly
consistent with the previous observations, the line effects are not
obvious in the $QU$ diagram. This is probably a result of the coarse
binning. Less coarse binning does not reveal any signatures as the
scatter increases considerably. Therefore, the results of M2007 are
used rather than these new data. Of the three remaining objects that
exhibit line effects, HK Ori and V586 Ori exhibit an excursion in $QU$
space and HD 179218 exhibits a clump of data surrounding the continuum
value, with a slight extension along the $U$ axis.

\smallskip

To arrive at the disk position angle from these spectropolarimetric
signatures, one should have some knowledge of the polarising
mechanism. In general, if the polarimetric signature is the result of
simple depolarisation, the polarisation vector in $QU$ space should be
measured from the line to the continuum. In case of intrinsic {\it
line} polarisation, the reverse is true, effectively resulting in a
difference of 90${\degr}$ from the former mechanism. To differentiate
between the two, the width of the line-effect is often used as a proxy
(see V2002). The intrinsic polarisation angles of HK Ori and V586 Ori
are determined via linear fits to their $QU$ excursions, assuming the
signature is due to depolarisation \citep[as the width of the
spectropolarimetric signatures is comparable to the emission line
width, see][]{Vink2002}. The polarisation angle of HD 179218 is
calculated assuming the excursion is only in the $U$ direction and
that the signature is due to intrinsic polarisation (since the
signature is narrower than the emission line). {\color{black}{{The uncertainties
in the polarisation angles calculated are approximately
$10^{\degr}$.}}}

\begin{center}
  \begin{figure*}
    \begin{center}

      \begin{tabular} {p{0.3\textwidth} p{0.3\textwidth} p{0.3\textwidth}}
	\includegraphics[width=0.3\textwidth]{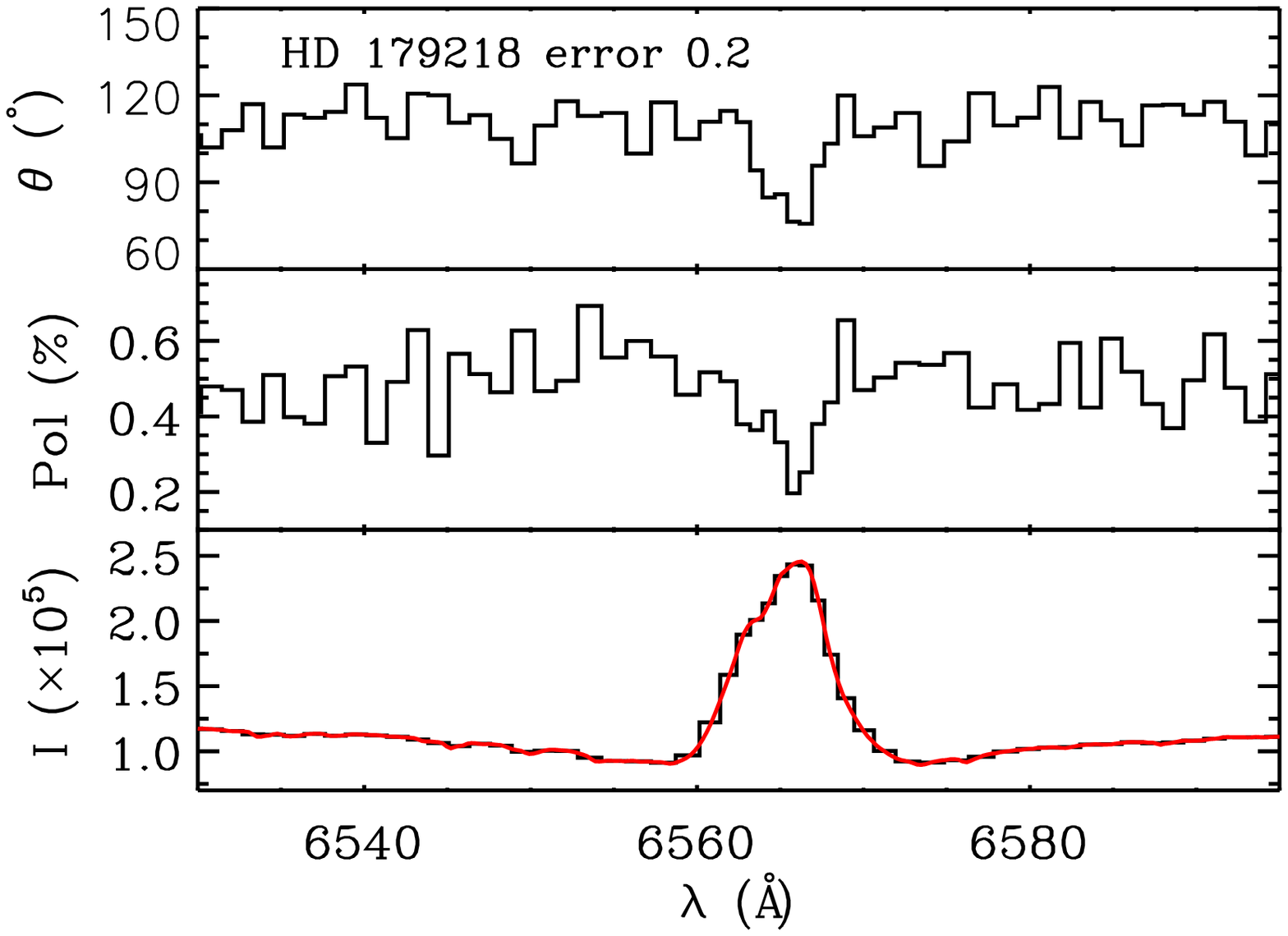} &
	\includegraphics[width=0.3\textwidth]{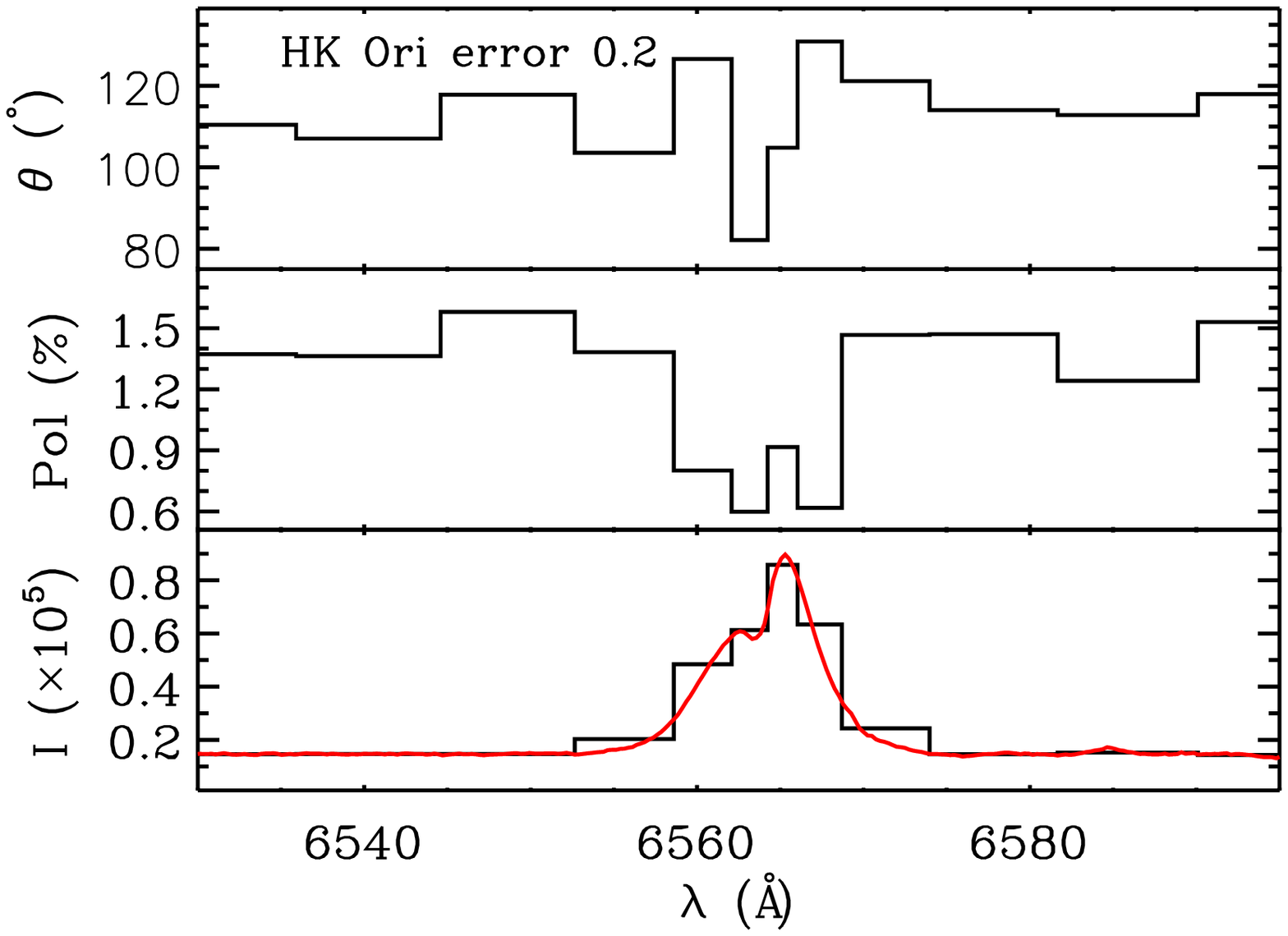} &
	\includegraphics[width=0.3\textwidth]{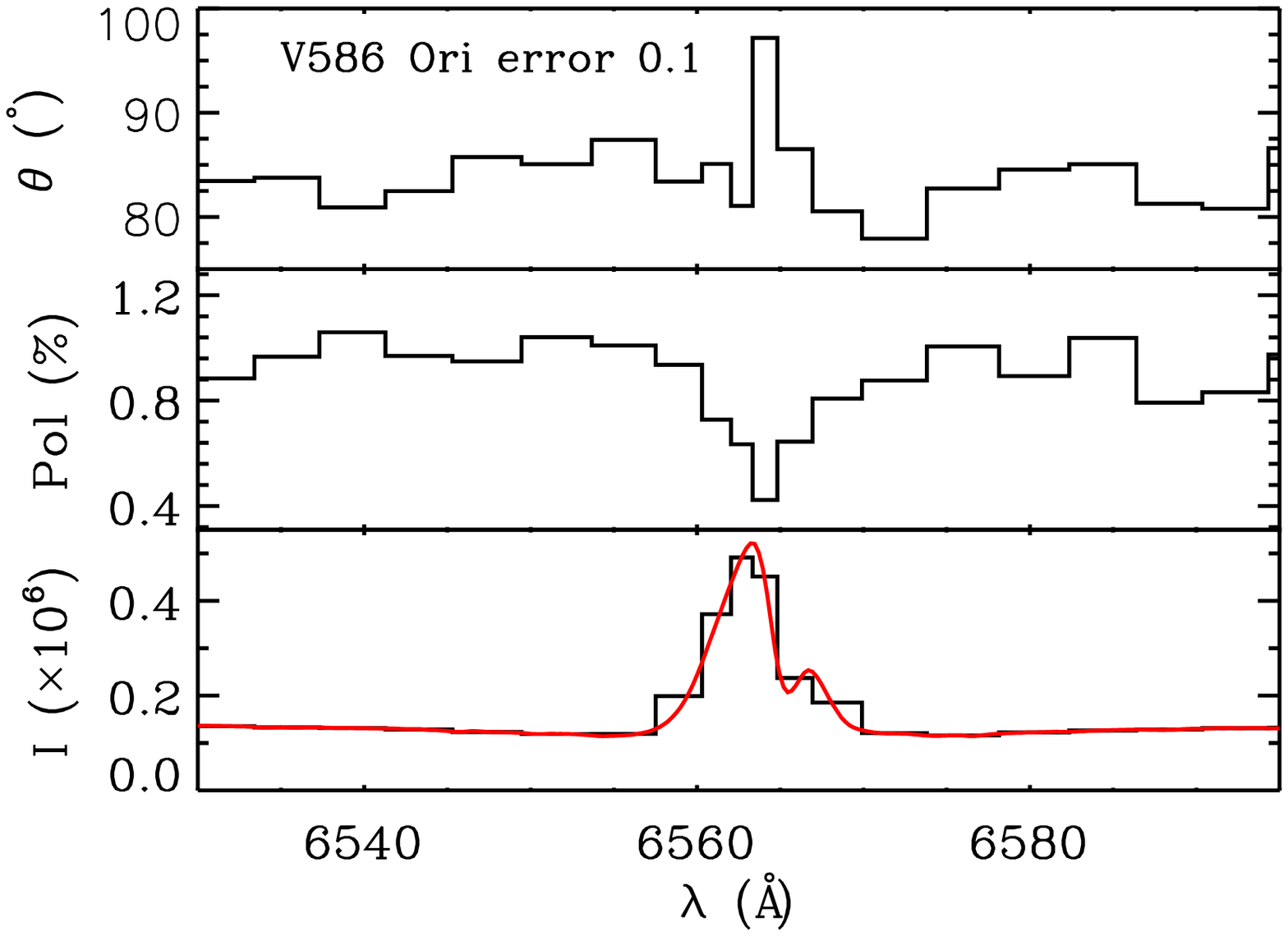} \\

	\includegraphics[width=0.3\textwidth]{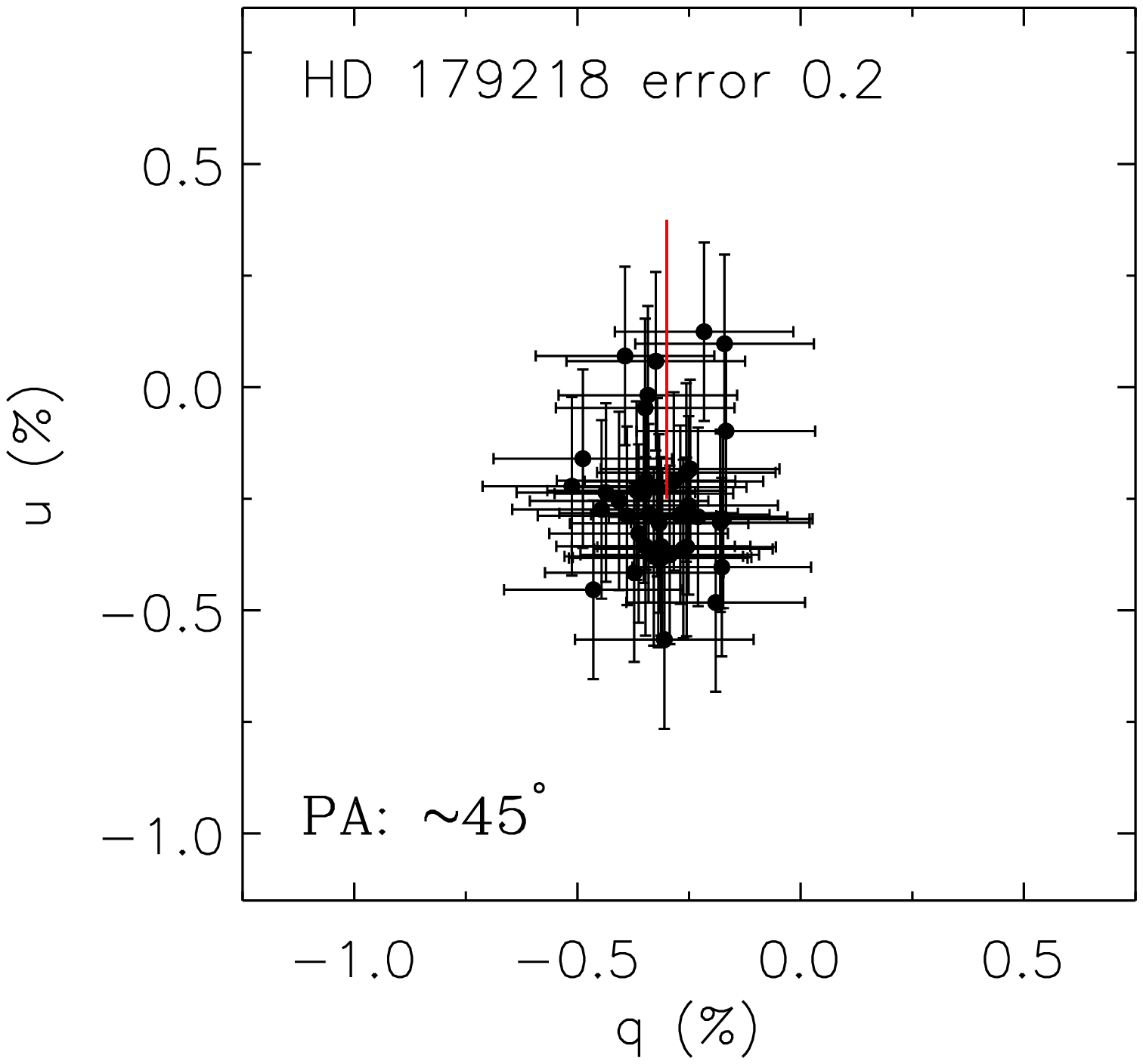} &
	\includegraphics[width=0.3\textwidth]{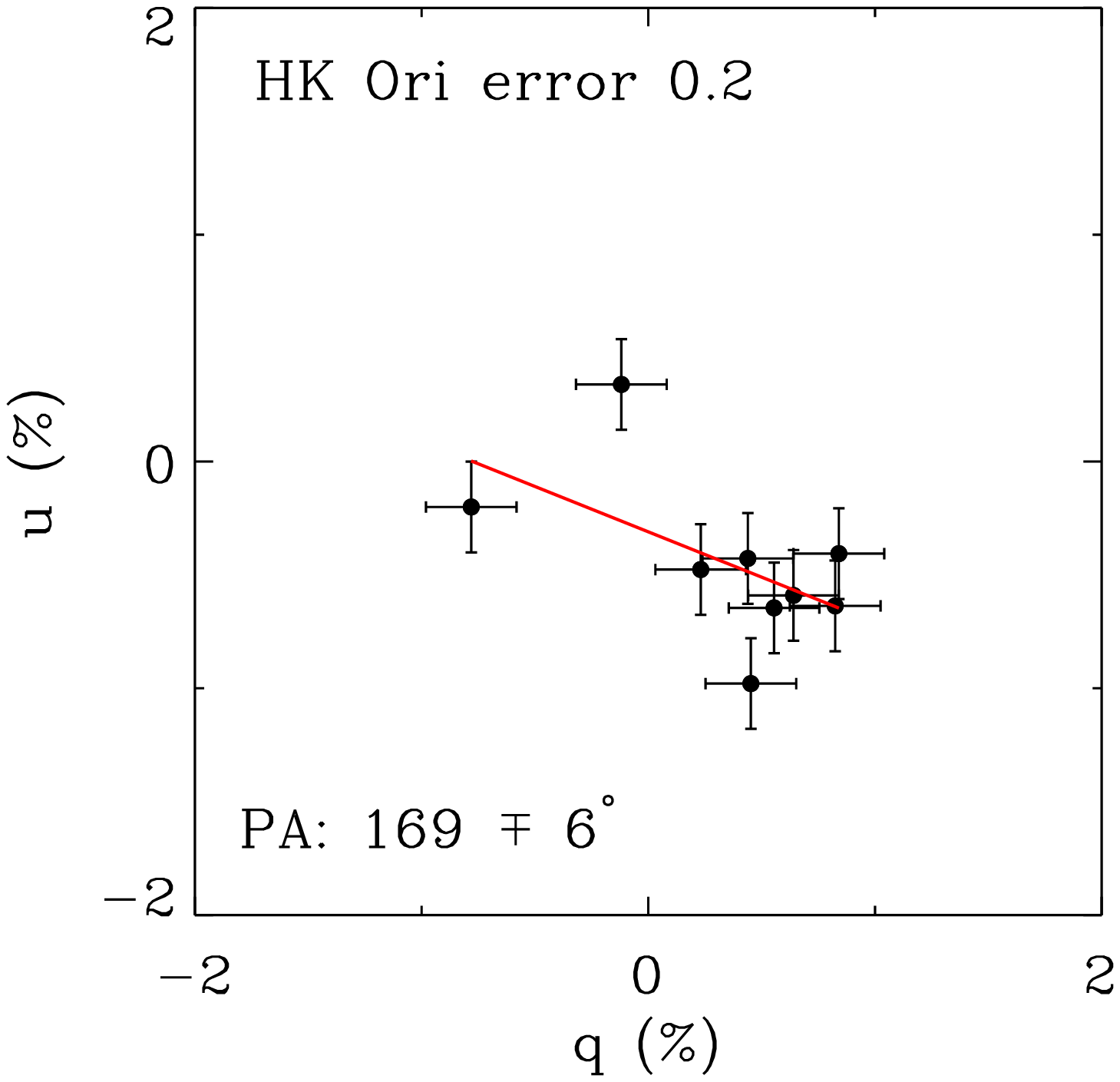} &
	\includegraphics[width=0.3\textwidth]{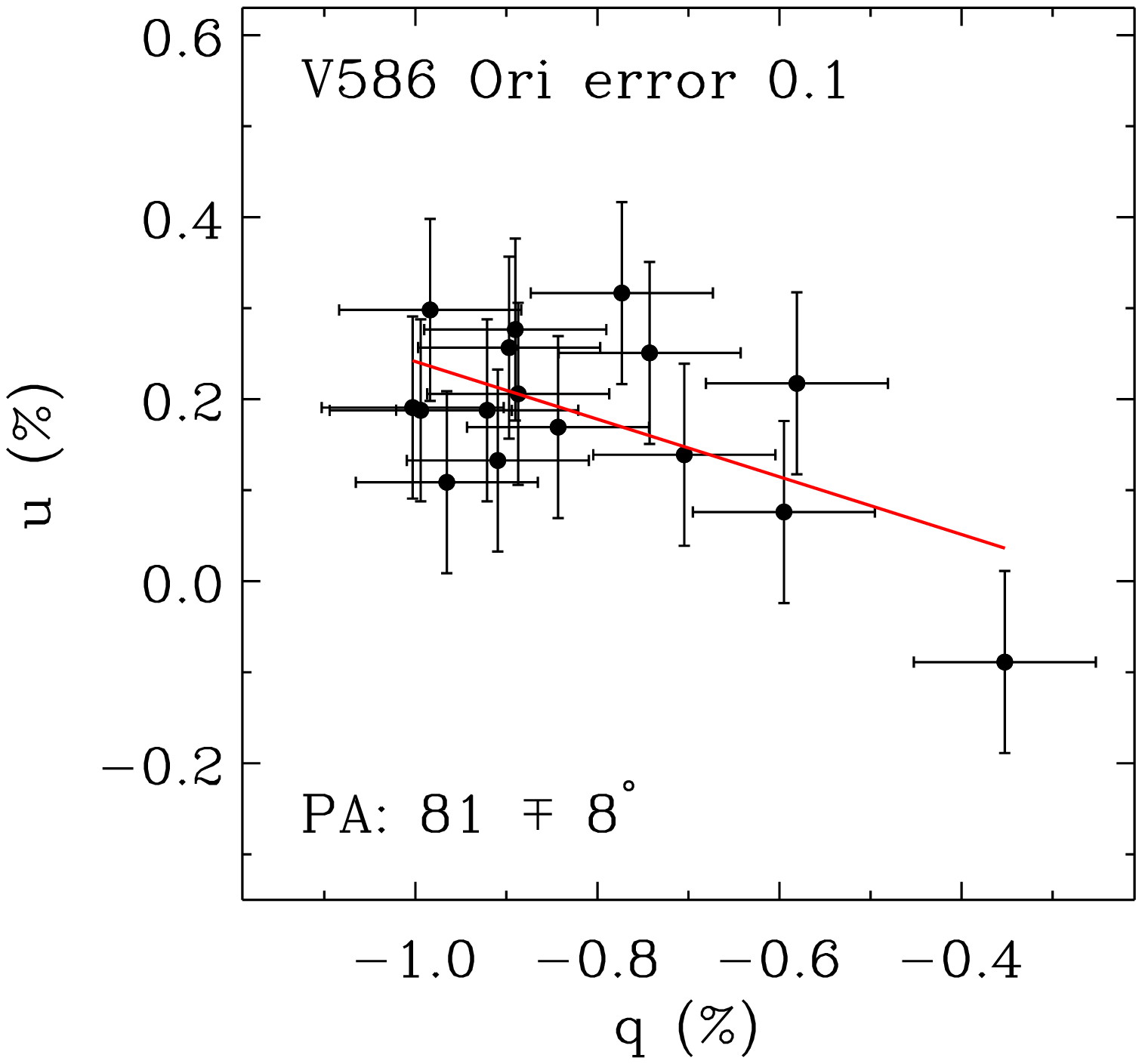} \\

      \end{tabular}

    \end{center}

    \caption{An example of the spectropolarimetric signatures of the
      sample. The top half of the figure presents the
      spectropolarimetric PA, the percentage polarisation, and the
      Stokes intensity spectra centred upon H$\alpha$. The
      data are binned to a constant {\color{black}{{polarisation}}} error, which is
      stated in the plots. The solid red line is the un-binned line
      profile. The $QU$ diagrams of the signatures are displayed in
      the lower half of the figure. The solid lines mark the direction
      of the intrinsic polarisation. \label{spec_pol_ex}}

  \end{figure*}
\end{center}

\smallskip 

{\color{black}{{The continuum polarisation presented in Table
\ref{spec_pol_obs} is broadly consistent with literature values
\citep[see e.g.][]{GMaheswar2002}. However, there are exceptions. For
example, we note that the continuum polarisation of HK Ori and MWC 758
differs from literature values. This may be due to intrinsic
variability as both objects are known to exhibit variable polarisation
\citep[see][]{DB2004,Besk1999}.}}}

\subsection{Literature data}

To supplement the spectropolarimetrically observed sample, intrinsic
polarisation angles (determined using spectropolarimetry) and binary
PAs were taken from the literature. The resultant sample of objects
for which both spectropolarimetric and binary PAs are available is
presented in Table \ref{data}. As the original studies from which the
data are taken were not primarily concerned with disk orientations, we
re-assessed all the polarisation angles taken from the literature to
ensure they are calculated consistently. The $QU$ data associated with
the PAs listed in Table \ref{data} were employed to determine
approximate angles. If these differed by more than
$\mathrm{90{\degr}}$ from the literature values, the reported angles
were rotated by $\mathrm{90{\degr}}$ (see the discussion above on
calculating disk PAs from spectropolarimetry). Table \ref{data}
presents the resultant sample.

The sample was then further supplemented by adding disk PAs determined
from direct imaging and multi-baseline interferometry. The additional
sample is also presented in Table \ref{data}.

\begin{center}
  \begin{table*}
    \begin{center}
	\caption{\label{data}Binary systems for which a measurement
	    of the intrinsic polarisation angle (from linear
	    spectropolarimetry) {\emph{or}} a direct constraint on the orientation of the circumprimary disk {\emph{and}} the PA of the binary system is
	    available. Column 3 denotes the spectral type of the
	    system primary, taken from SIMBAD unless otherwise
	    stated. Column 4 lists the binary PA, column 5 lists the intrinsic polarisation
	    angle and column 6 contains the disk PA. Given the typical errors, the angles are presented to the closest degree.}

	\begin{tabular}{l l l r r r}
	  \hline
	  Object & Alt. Name & Type &   \multicolumn{1}{c}{Bin. PA} &  \multicolumn{1}{c}{Pol. PA} & \multicolumn{1}{c}{Disk PA}\\
	  &          &   & \multicolumn{1}{c}{($\mathrm{{\degr}}$)} &   \multicolumn{1}{c}{($\mathrm{{\degr}}$)} & \multicolumn{1}{c}{($\mathrm{{\degr}}$)}\\
	  \hline
	  \hline
\multicolumn{5}{l}{Only spectropolarimetry}\\
	  \object{MWC 166}     & HD 53367                         & B0 & 298\mkref{1} &  46\mkref{2} &  \\
	  \object{HD 58647}    & BD $-$13$\mathrm{{\degr}}$ 3008 & B9 & 115\mkref{3} & 20\mkref{4}  &  \\
	  \object{MWC 158}     & HD 50138                         & B9 & 30\mkref{3}    & 135\mkref{4} &\\
	  \object{MWC 120}     & HD 37806                         & A2 & 34\mkref{1}  & 90\mkref{4}  & \\
	  \object{V586 Ori}    & HD 37258                         & A2 & 217\mkref{1} &  81\mkref{5} &  \\
	  \object{T Ori}       & MWC 763                          & A3 & 107\mkref{1} & 20\mkref{4}  &    \\
	  \object{HK Ori}      & MWC 497                          & A4 & 47 \mkref{1} & 169\mkref{5} & \\

	  \hline

	  \multicolumn{5}{l}{Spectropolarimetry and imaging/interferometry}\\
	 
      \object{HD 200775}                      & MWC 361     & B2                      & 164\mkref{6} & 93\mkref{2}       & 7\mkref{7}   \\        
      \object{MWC 147}                        & V700 Mon    & B6                      & 82\mkref{1}   & 168\mkref{2}      & 80\mkref{8}    \\
      \object{HD 45677}                       & FS CMa      & B2\mkref{9}             & 150\mkref{3}    & 164\mkref{10}      & 77\mkref{11}    \\
      BD +40$\mathrm{{\degr}}$ 4124 & \object{V1685 Cyg}   & B3                      & 175\mkref{12}   & 36\mkref{2}       &  110\mkref{13}   \\
      \object{MWC 1080}                       & V628 Cas    & B0                      & 269\mkref{1}  & 75\mkref{2}       & 55\mkref{13}     \\
      \object{CQ Tau}                         & HD 36910    & F3                      & 56\mkref{12}   & 20\mkref{4}       & 120\mkref{14}\\
      \object{HD 179218}                      & MWC 614     & A0                      & 141\mkref{12}   & $\sim$45\mkref{5} & 23\mkref{15}    \\

\hline

\multicolumn{5}{l}{Only imaging/interferometry}\\
      \object{MWC 758}   & HD 36112      & A5   & 311\mkref{12}     &  & 128\mkref{13}     \\
      \object{V892 Tau}  & HBC 373       & B8  & 23\mkref{12}       &  & 53\mkref{16}        \\
      \object{R Mon}     & MWC 151       & B0  & 287\mkref{17}        &  & $\sim$80\mkref{18}  \\
      \object{MWC 297}   & NZ Ser        & B0  & 313\mkref{19}        &  & 165\mkref{20}         \\
      \object{HR 5999}   & V856 Sco      & A7 & $\sim$111\mkref{21} &  & $\sim$25\mkref{22}   \\
      \object{HD 101412} & PDS 57        & B9.5 & 226\mkref{12}     &  & 38\mkref{15}         \\

      \hline	
    \end{tabular}

    \tablebib{1: From the data presented in \citet{Wheelwright2010}, 2: M2007, 3: \citet{DB2006}, 4: \citet{Vink2005a}, 5: These data, see Section \ref{sigs}, 6: \citet{Pirzkal1997}, 7: \citet{Okamoto2009}, 8: \citet{Kraus2008}, 9: \citet{Cidale2001}, 10: \citet{Patel2006}, 11: \citet{Monnier2006}, 12: \citet{Thomas2007}, 13: \citet{Eisner2004}, 14: \citet{Doucet2006}, 15: \citet{Fedele2008}, 16: \citet{Monnier2008}, 17: \citet{Weigelt2002}, 18: \citet{Fuente2006}, 19: \citet{Vink2005b}, 20: \citet{Manoj2007}, 21: \citet{Stecklum1995}, 22: \citet{Preibisch2006}.}

  \end{center}
  \end{table*}
\end{center}

\subsection{Modelling co-planar binary systems and disks}

\label{model}

In order to compare disk and binary PAs in a meaningful way, a model
is required to predict the difference in the two angles expected for
various scenarios. \citet{DB2006} simply compare disk and binary
PAs. However, binary PAs do not necessarily relate to the binary
orbital plane. For example, even when a binary system and its
circumprimary disk lie in the same plane, if the system is seen
face-on, the binary PA is unrelated to the PA of the disk. In
the case of more edge-on systems, the binary PA is likely to be
aligned with the disk PA. Nonetheless, at certain phases of the orbit,
the binary PA will be quite different to the circumprimary disk
PA. It is only in the extreme case of an edge-on, co-planar system
that the disk and binary PAs are constantly aligned. Here we employ a
new model to predict the average distribution in the
difference between disk and binary PA when circumstellar disks and
binary orbits lie in the same plane.

\smallskip

The model is characterised by a random orbital phase; inclination;
semi-major axis and PA of the line about which the system is
inclined. The eccentricity of the system was kept constant. Although a
constant eccentricity is not truly representative of the eccentricity
distribution of PMS binary systems \citep[see e.g.][]{Goodwin2007},
neither is a random eccentricity distribution. Since the eccentricity
can affect the results, the input eccentricity is left as a constant
and is treated as a free parameter. {\color{black}{{This allows us to fit the
observed distribution in differences between binary and disk PAs}}}. The
masses of the components were kept constant at 6 \&
1~${M}_{\odot}$. The component masses have little influence on
the final distribution and are mentioned only for completeness.

\smallskip

We used 10$\,$000 random systems to determine the distribution in the
difference between the disk and binary PAs
($\Delta$PA). Spectropolarimetry is insensitive to face on systems as
the projected polarisation vectors cancel one-another out. Therefore,
systems with low inclinations are discarded. The value of the cut-off
inclination does {\color{black}{{affect}}} the final
distribution. However, so long as large values
(e.g.$\mathrm{>50{\degr}}$) are not chosen, the differences between
the final distributions are relatively small. Since we also use
imaging observations which are less sensitive to inclination, a low
cut-off value ($\mathrm{10{\degr}}$) is used. {\color{black}{{The
imaging and spectropolarimetric samples have different cut-off
values. However, neither sample is large enough to be used in
isolation, and thus a single cut-off value is required. Changing the
cut-off value has a similar effect to changing the eccentricity. Since
we use the eccentricity as a free parameter, the use of a single
cut-off value for the sample of both imaging and spectropolarimetric
data will not prevent a fit to the data. By varying the eccentricity
to fit the data, the uncertainty in the cut-off value is essentially
incorporated into the best-fitting eccentricity as a systematic
uncertainty.}}}

\smallskip

\begin{center}
  \begin{figure}
    \begin{center}

      \includegraphics[width=0.45\textwidth]{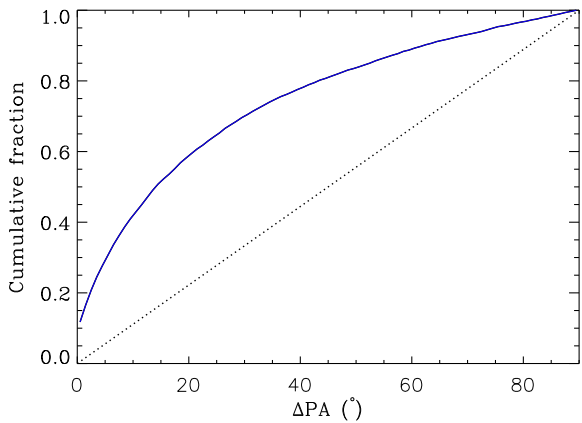}

      \caption{\label{a_test}The expected distribution in the
	difference between position angles of disks and binary systems
	in a co-planar model ({blue} solid line) {\color{black}{{with an eccentricity
	of $e=0.5$}}}. The black dotted line represents the expected
	distribution due to a random orientation of binary and disk
	PAs.  }
    \end{center}
  \end{figure}
\end{center}

Figure \ref{a_test} presents a typical distribution. We are only
concerned with the magnitude of the difference between disk and binary
PAs. Therefore, we convert all differences between these angles to be
in the range $\mathrm{0-90\degr}$. For example, a difference of
$\mathrm{170\degr}$ is $\mathrm{10\degr}$ away from alignment, and is
thus classified as a difference of $\mathrm{10\degr}$. An alignment
between disk and binary PAs corresponds to an offset of
0$\mathrm{{\degr}}$. It can be seen that the distribution tends
towards intrinsic alignment, and appears noticeably different to the
distribution expected if the two angles are not related. Specifically,
the distribution does not exhibit a direct correlation between binary
and disk PA, but does demonstrate an excess of aligned angles over the
random distribution. This is to be expected based on the previous
discussion. To reiterate, co-planar disks and orbits are more likely
to be observed to be aligned than not, since at high inclinations the
disk and binary PAs will be the same. This is not the case for
non-coplanar systems. Therefore, the co-planar distribution exhibits a
preference for an alignment between disk and binary position angles
compared to the random distribution.

{\color{black}{{We note that comparing this model to observations
assumes that the orientation of the circumstellar disks in the sample
is primordial. This need not be the case. Even if the circumstellar
disks in HAe/Be binary systems originally lie in the plane of the
binary orbits, gravitational interactions with another star have the
potential to alter their orientation \citep[see e.g. the discussion
in][]{Bate2000}. This would weaken an intrinsic correlation between
binary and disk PA. Therefore, any observed correlation between disk
and binary PA may be more significant than it initially
appears. However, quantifying this is beyond the scope of the
paper.}}}

\subsection{The relationship between binary and disk position angle}

\label{alignment}

\begin{center}
  \begin{figure}
    \begin{center}
      \begin{tabular}{c}
	\includegraphics[width=0.45\textwidth]{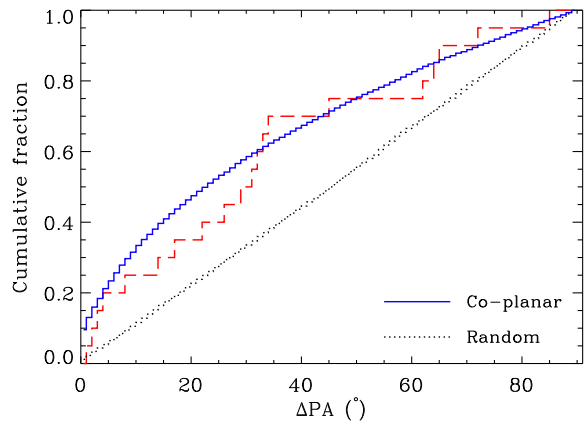} \\
      \end{tabular}

      \caption{\label{res3}The difference in disk and binary PAs
presented in Table \ref{data} (red dashed). This is compared to the co-planar model with a constant
eccentricity of 0.1 (blue solid) and a distribution generated by
assuming the two angles are randomly aligned (black dotted). Note
that $\Delta$PA refers to the difference between binary and disk
position angle (i.e. accounting for the 90$\mathrm{{\degr}}$ offset
between spectropolarimetric signatures and disks).}

    \end{center}
  \end{figure}
\end{center}

The final sample of HAe/Be binary systems with disk PAs contains 20
objects, and is thus more than three times greater than that of
\citet{DB2006}. The differences between disk and binary PAs in the
sample are compared to the co-planar model and a random
distribution. In Figure \ref{res3} we show the cumulative distribution
of the differences in PA between the binary systems and those derived
for the disks. Also in the figure are the distributions predicted by
the co-planar model and a completely random association of disk and
binary PAs. A first glance indicates that the aligned scenario
provides a better fit to the data than the random distribution. This
is corroborated by a Kolmogorov-Smirnov (KS) test; while the co-planar
model is found to be consistent with the data (it can only be
ruled out at the 0.2$\sigma$ level), the random hypothesis is barely
consistent with the data, and can be rejected at a level of
2.2$\sigma$.

\smallskip

This 2$\sigma$ level of rejection may seem low, but we note that both
models are more similar in their cumulative distributions than one
might naively expect. As mentioned earlier, this is because inclined
binary systems do not necessarily have the same observed orientation
as their circumprimary disks.  This may explain why our significance,
with an improved data-set, is of the same order as \citet{DB2006} found
with a smaller data-set and a more simplistic model.

\smallskip

Improving the current statistics requires further enlarging the
sample. We find that if the data followed the model distribution
exactly, differentiating between the two scenarios at a 3$\sigma$
level would require a sample of approximately 50 objects. Nonetheless,
the current data are consistent with the co-planar hypothesis and
furthermore, this scenario is favoured over random alignments.

\section{Does spectropolarimetry really trace circumstellar disks?}
\label{initial_test}

Our use of spectropolarimetry assumes that the spectropolarimetric
signatures of HAe/Be stars can be used to trace the orientations of
their circumstellar disks. {\color{black}Linear polarisation has been
used for some time to probe the orientation of circumstellar disks
\citep[see e.g.][]{Bastien1990}}. However, the proposal of
\citet{Harrington2007} and \citet{Kuhn2007} that the
spectropolarimetric signatures of HAe/Be stars are due to optical
pumping and not scattering in a disk implies that these signatures do
not have to trace disks. We address this issue here and do this by
comparing spectropolarimetric observations of young stellar objects
with independent measurements of the orientation of their disks. To
increase the sample we include observations of both HAe/Be and T Tauri
stars.

\smallskip

Table \ref{sample_spec_pol} presents the sample of young stellar
objects, both low mass T Tauri objects and Herbig Ae/Be stars, for
which the combination of spectropolarimetric observations and
independent observations of their circumstellar disks is available. We
first concentrate on the HAe/Be sample and then assess whether there
is a general trend within the combined sample of young stellar
objects. The majority of HAe/Be disks are thought to be optically thin
\citep[see e.g.][]{Natta2001}. If the observed polarisation signatures
are due to single scattering in these disks, the polarisation angles
will be perpendicular to the disks' PA on the sky. Therefore, if the
spectropolarimetric signatures of HAe/Be stars are due to disks, the
difference between the disk and polarisation angles for this sample
will be $\mathrm{90{\degr}}$. {\color{black}{{In testing this
hypothesis, we allow for an uncertainty in the difference between
polarisation and disk PAs of $\sim$15${\degr}$, which is typical for
the values used. In some cases, there is a systematic uncertainty due
to different PAs being reported by different authors. For example, in the
case of AB Aur, \citet{Mannings1997} report a disk PA of
$\mathrm{79^{\degr}}$ while \citet{Corder2005} report various possible
angles ranging from $\mathrm{26^{\degr}}$ to
$\mathrm{85^{\degr}}$. However, such cases are not common, and are
unlikely to influence conclusions regarding the whole sample.}}}

\begin{center}
  \begin{table*}
    \begin{center}
      
      \caption{Young stellar objects (column 1) for which
      spectropolarimetric observations \emph{and} a direct measurement
      of their disk PA are available. The disk and adopted polarisation
      angles are presented in columns 4 and 5 and the difference between
      them is listed in column 6. Finally, column 7 indicates objects
      where the difference is close to $\mathrm{90{\degr}}$ ($\mathrm{\perp}$) or
      $\mathrm{0{\degr}}$ ($\mathrm{\parallel}$).\label{sample_spec_pol}}

      \begin{tabular}{l l l r r r c}
	\hline
	Object & Alt. Name & Type & \multicolumn{1}{c}{Disk PA} &  \multicolumn{1}{c}{Pol. PA} & \multicolumn{1}{c}{$\Delta$PA} & \\
	&           &  & \multicolumn{1}{c}{($\mathrm{\degr}$)}  & \multicolumn{1}{c}{($\mathrm{\degr}$)}& \multicolumn{1}{c}{($\mathrm{\degr}$)} &\\
	
	\hline
	\hline
	\multicolumn{4}{l}{\bf{HAe/Be}}\\

	\object{HD 200775}   & MWC 361   & B2          & 7\mkref{1}            &  93\mkref{2}       & 86 & $\mathrm{\perp}$\\ 
	\object{MWC 147}     & V700 Mon  & B6          & 80\mkref{3}             &  168\mkref{2}      & 88 & $\mathrm{\perp}$\\
	\object{HD 45677}    & FS CMa    & B2\mkref{4} & 77\mkref{5}             &  164\mkref{6}      & 87 & $\mathrm{\perp}$\\
	BD +40$\mathrm{{\degr}}$ 4124 & \object{V1685 Cyg} & B3          & $\mathrm{110}$\mkref{7} &  36\mkref{2}       & 74 & $\mathrm{\perp}$\\
	\object{MWC 1080}    & V628 Cas  & B0          & $\mathrm{55}$\mkref{7}  &  75\mkref{2}       & 20 & $\mathrm{\parallel}$\\
	\object{CQ Tau}      & HD 36910  & F3 & 120\mkref{8}            &  20\mkref{9}      & 80 & $\mathrm{\perp}$\\
	\object{MWC 480}     & HD 31648  & A3          & $\mathrm{150}$\mkref{7} &  55\mkref{9}      & 85 & $\mathrm{\perp}$\\
	\object{AB Aur}      & HD 31293  & A0          & $\mathrm{79}$\mkref{10} &  160\mkref{9}     & 81 & $\mathrm{\perp}$\\
	\object{HD 179218}   & MWC 614   & A0IVe       & 23\mkref{11}            & $\sim$45\mkref{12} & 22 & $\mathrm{\parallel}$\\

	\multicolumn{4}{l}{\bf{T Tauri}}\\

	\object{RY Tau}      & HD 283571 & F8          & 62\mkref{13}            & 163\mkref{9}      & 79 & $\mathrm{\perp}$\\
	\object{SU Aur}      & HD 282624 & G2          & 127\mkref{14}            & 130\mkref{9}      & 3  & $\mathrm{\parallel}$\\
	\object{FU Ori}      & HBC 186   & G3          & 47\mkref{15}            & 45\mkref{9}       & 2  & $\mathrm{\parallel}$\\
	\object{GW Ori}      & HD 244138 & G5          & 56\mkref{16}            & (60)\mkref{9}     & 4  & $\mathrm{\parallel}$\\
	\object{DR Tau}      & HBC 74    & K5          & 128\mkref{17}           & 120\mkref{9}      & 8  & $\mathrm{\parallel}$\\

	\hline
      \end{tabular}
      \tablebib{1: \citet{Okamoto2009}, 2: M2007, 3:
	\citet{Kraus2008}, 4: \citet{Cidale2001}, 5:
	\citet{Monnier2006}, 6: \citet{Patel2006}, 7:
	\citet{Eisner2004}, 8:
	\citet{Doucet2006}, 9: \citet{Vink2005a}, 10: \citet{Mannings1997}, 11:
	\citet{Fedele2008}, 12: these data, 13: \citet{Akeson2003},
	14: \citet{Akeson2002}, 15: \citet{Malbet2005}, 16:
	\citet{Mathieu1995}, 17: \citet{Kitamura2002}.}
    \end{center}
  \end{table*}
\end{center}

\label{test}

\subsection{Herbig Ae/Be stars}

In the case of the HAe/Be stars in Table \ref{sample_spec_pol}, the
majority of the objects (7 out of 9) have polarisation angles
approximately perpendicular to their imaged disks. This might be
expected if the polarisation signatures are due to single scattering in
disks. Therefore, this appears to validate the use of
spectropolarimetry to trace disks. Here we quantify this. We use the
HAe/Be star data in Table~\ref{sample_spec_pol} to test the hypothesis
that the intrinsic polarisation angle is always perpendicular to the
disk PA (within the errors). The cumulative distribution of the
difference in disk PA and polarisation PA is shown in Figure
\ref{new_hyp}. We find that the hypothesis that the disk and
polarisation angles are unrelated to each other and thus randomly
oriented can be discounted at a significant level (at 3.1$\sigma$
according to the KS test). In contrast, the hypothesis that the
spectropolarimetric signatures of the HAe/Be stars are oriented
perpendicularly to their disks cannot be rejected beyond the 1$\sigma$
level, and is thus consistent with the data. Therefore, we find that
spectropolarimetric signatures of HAe/Be stars do trace the
orientation of their circumstellar disks.

\smallskip

We note that, although the majority of HAe/Be objects exhibit a
difference in disk and polarisation angle that is close to
$\mathrm{90{\degr}}$, two objects have disk and polarisation angles
that are essentially aligned. This is contrary to expectations based
on single scattering occurring in an optically thin disk. In their
smaller sample, \citet{Vink2005a} also note several objects where this is the
case. They suggest that while the spectropolarimetric signatures of
all young stellar objects are due to circumstellar disks, the angle of
the resulting polarisation vector is dependent upon the properties of
the inner disk. If the inner disk is optically thin, single scattering
dominates and the resulting polarisation vector is perpendicular to
the disk PA. In contrast, if the inner disk is optically thick, the
polarisation vector is parallel to the disk PA due to multiple
scattering.

\smallskip

Many of the T Tauri stars in Table \ref{sample_spec_pol} also have
disk and polarimetric PAs that are aligned. Therefore, if this
argument is correct, it would appear that the majority of T Tauri star
disks are optically thick in the inner regions.

\subsection{Removing the dependence on optical depth}

Although the previous test provides a statistically significant
result, it could be argued that this is dependent upon prior knowledge
of the disks' optical depth. Assuming HAe/Be star disks are optically
thin is in part justified. The vast majority of the HAe/Be stars
considered have an offset between disk and polarisation angle that is
close to 90$\mathrm{{\degr}}$, as expected for single scattering in optically
thin disks. However, it could be argued that this is a circular
argument. Moreover, in the case of the T Tauri objects, the disk and
polarisation angles are generally aligned. A more appropriate
hypothesis for the total sample may be a combination of the two
scenarios, i.e. that polarisation angles {\emph{are}} related to the
PAs of disks, but that the polarisation angles can be either aligned
or perpendicular to the disk. In general, we do not have prior
knowledge of the optical depth of the disks. Therefore, we test the
relationship between polarisation and disk angle without making
{\emph{a priori}} assumptions.

\smallskip

\begin{center}
  \begin{figure*}
    \begin{center}
\begin{tabular}{l l}
        \includegraphics[width=0.45\textwidth]{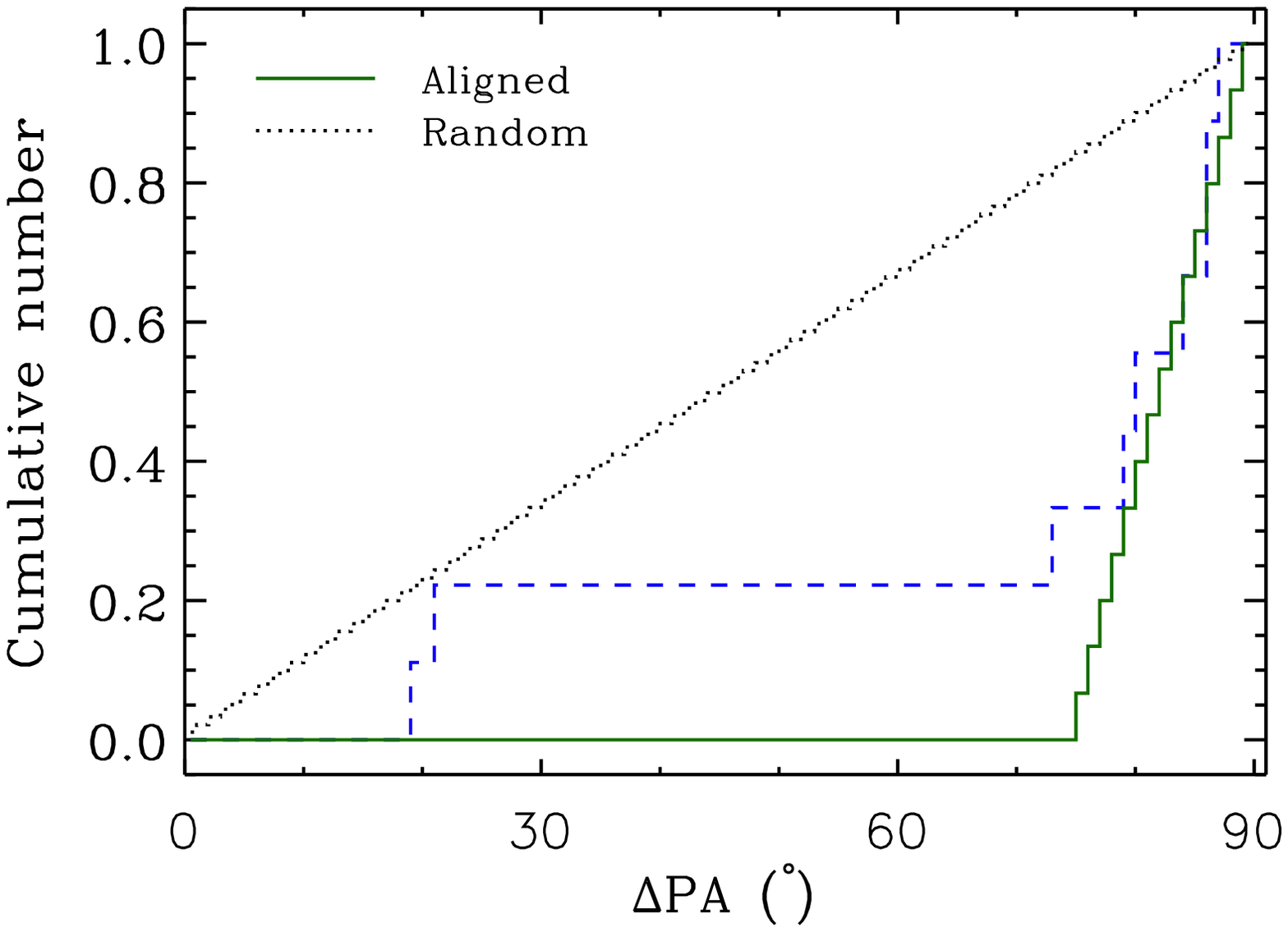} &
      \includegraphics[width=0.45\textwidth]{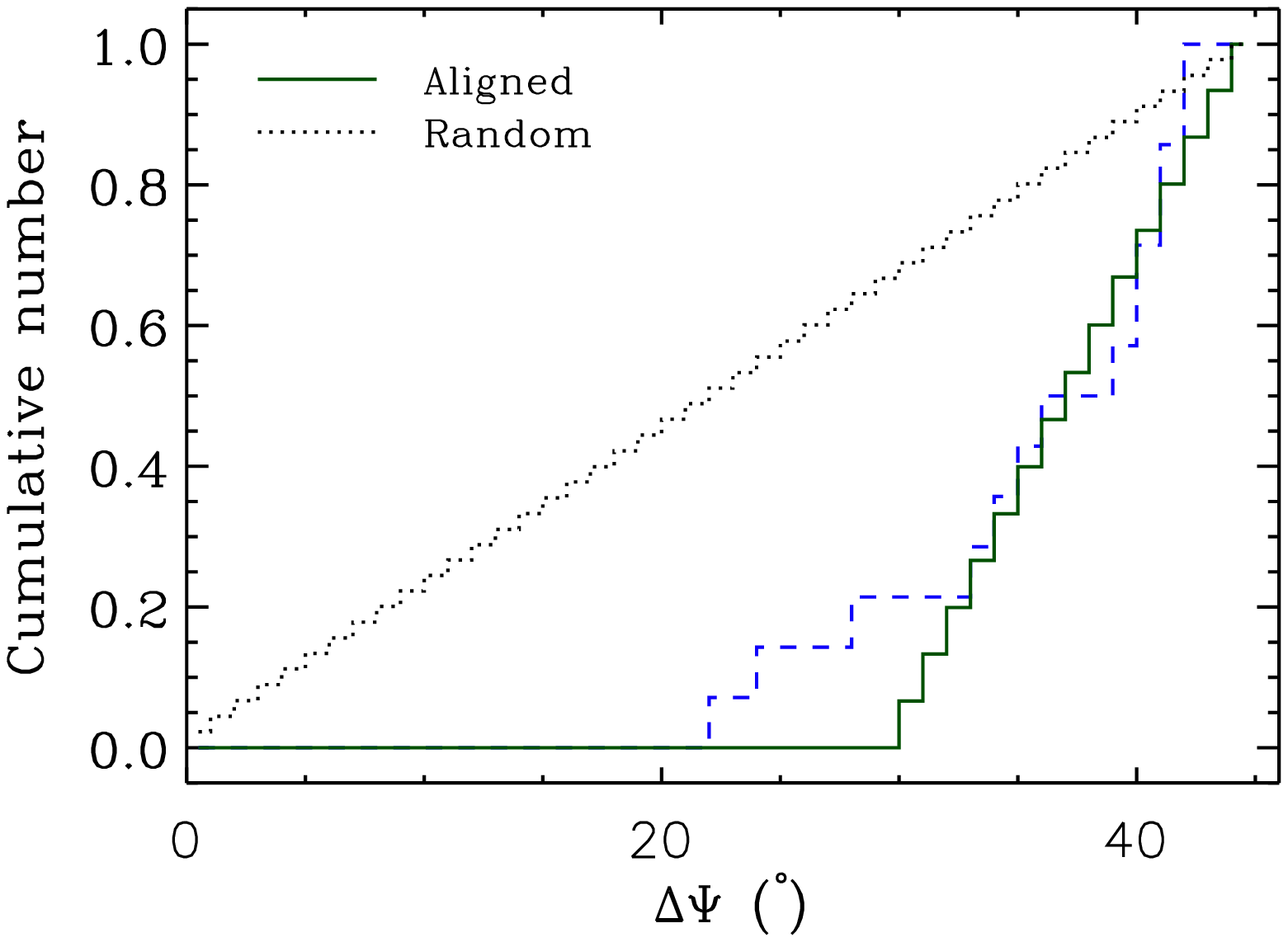}\\
\end{tabular}

      \caption{The distribution in the difference between
spectropolarimetrically predicted disk PA and observed disk PA for the
sample presented in Table \ref{sample_spec_pol} (blue dashed). This is
compared to a random distribution (black short dotted). On the left we show a distribution
where polarisation signatures are always orientated perpendicularly to
circumstellar disks and on the right we present the distribution for a scenario where the
spectropolarimetric signatures can be either perpendicular or parallel
to disks (see the text for more detail). Both model distributions have
a maximum error of $\mathrm{15{\degr}}$. In both cases, a random
orientation of disk and polarisation position angles can be discarded at the 3$\sigma$ level.\label{new_hyp}}
    \end{center}
  \end{figure*}
\end{center}

In the scenarios mentioned above, the difference between the disk and
spectropolarimetric angle is either 0 or 90${\degr}$, depending upon
the optical depth of the inner disk. Consequently, it can be expected
that the offset from 45${\degr}$ to the difference between disk and
spectropolarimetric angle (henceforth $\Delta \Psi$) is always
45$\mathrm{\degr}$. We note that this is the case regardless of
whether the signature is interpreted as being due to line polarisation
or depolarisation, or whether the disk is optically thin or
thick. Therefore, this test is even more robust than the previous
which assumes optically thin scattering and is subject to the
polarisation signature being interpreted correctly. Here we compare
this hypothesis to the sample presented in Table
\ref{sample_spec_pol}. The disk and spectropolarimetric angles in
Table \ref{sample_spec_pol} are used to calculate $\Delta \Psi$. This
is then compared to the hypothesis that $\Delta \Psi$ is 45${\degr}$
by calculating the average of 10$\,$000 equally sized samples in which
$\Delta \Psi$ is 45${\degr}$ but with an additional random error contribution
with a maximum value of 15$\mathrm{{\degr}}$(see Figure
\ref{new_hyp}).

The hypothesis that $\Delta \Psi$ is random can be discounted at a
significant level (above 3$\sigma$). This leaves the hypothesis that
$\Delta \Psi$ is always 45$\mathrm{{\degr}}$, which cannot be rejected
at greater than a 1$\sigma$ level and is thus consistent with the
data. This confirms the earlier finding that spectropolarimetric
signatures trace disks and demonstrates that this is not dependent on
spectral type and classification (i.e. T Tauri star or HAe/Be star).

\smallskip

To summarise, we show that the spectropolarimetric data presented in
Table \ref{sample_spec_pol} do appear to trace circumstellar
disks. This would be expected if the polarisation is due to scattering
in these disks. However, \citet{Harrington2009a} claim many of the
HAe/Be stars in our sample have polarisation signatures which require
optical pumping and absorption in outflows (e.g. AB Aur and MWC
480). We show here that regardless of the polarising mechanism,
spectropolarimetry can be employed to trace the orientation of
circumstellar disks.

\section{Discussion and Conclusion}
\label{disc}

\subsection{On the use of spectropolarimetry to probe circumstellar disks}

The results presented in this paper indicate a direct correlation
between the spectropolarimetric signatures of pre-main-sequence stars
and the orientations of their circumstellar disks. This is significant
above the 3$\sigma$ level and appears independent of the
classification of the young stellar objects. Therefore, we conclude
that spectropolarimetric signatures of young stellar objects do indeed
trace the orientation of their circumstellar disks. This is expected
in the case of polarisation of stellar and accretion shock photons by
disks \citep{McLean1979,Vink2002}. In the case of polarisation via
optical pumping and absorption, the relationship between
spectropolarimetric signatures and disks is less clear.

\smallskip

We note that \citet{Kuhn2010} suggest that the polarisation signature
of the Herbig Be star HD 200775 is due to optical pumping and that the
signature does trace the orientation of an imaged disk. However, the
spectropolarimetric signature of this object is observed across a
double-peaked emission line profile, and might therefore be the result
of depolarisation after all. Nevertheless, many Herbig Ae/Be stars,
several of which are in our sample, exhibit polarisation signatures
associated with P~Cygni line profiles, i.e. outflowing gas \citep[see
e.g.][]{Harrington2009a}. Here we show that the
signatures still appear to trace the orientation of circumstellar
disks. This implies that, if the spectropolarimetric signatures are
due to optical pumping and absorption in a wind, the wind geometry
essentially mirrors that of the disk, at least in the regions where
the polarisation occurs.

\smallskip

This is partly substantiated by recent observations of the H$\alpha$
emission of the Herbig Ae star AB Aur with the CHARA array by
\citet{Perraut2010}. This object has been proposed to exhibit
polarisation due to optical pumping since it displays polarisation
across the P~Cygni absorption component of its H$\alpha$ emission
\citep{Harrington2007}. \citet{Perraut2010} resolved the H$\alpha$
emitting region around AB Aur and found that it could be modelled as
the base of a wind represented by a flattened torus encompassing a
circumstellar disk. Provided the inclination and the angle between the
wind surface and disk mid-plane is low \citep[$\mathrm{20{\degr}}$ and
35$\mathrm{{\degr}}$ in the case of the disk-wind model
of][]{Perraut2010}, such a flattened torus might well appear to have a
similar morphology to the disk.

\subsection{On the alignment between binary systems and their circumprimary disks}

To investigate the relative alignment of HAe/Be binary systems and
circumstellar disks we have used spectropolarimetry and high spatial
resolution data to determine the orientation of circumstellar disks
around the primary components of such systems. We then combined these
disk angles with binary parameters to assess whether HAe/Be
circumstellar disks and binary systems are
co-planar. {\color{black}{{Studies of lower mass T Tauri stars have
found that the circumstellar disks in T Tauri star binary systems tend
to be aligned, suggesting that such systems may form via fragmentation
\citep[see e.g.][]{SWolf2001,Jensen2004,Monin2006}. Here we
investigate whether this is also the case for HAe/Be systems.}}} We
note that \citet{GMaheswar2002} also compared HAe/Be binary and
polarisation angles (although these were calculated via broadband
polarimetry and thus subject to uncertainties in the correction for
the interstellar polarisation). These authors find that their data are
inconsistent with a random association of disk and binary position
angles with a significance of 84 per cent. This is not very conclusive
and the authors note that they do not account for projection effects
and thus the actual correlation may be stronger. We do account for
projection effects and demonstrate that, in principle, co-planar and
randomly orientated disks and binaries can be differentiated.

\smallskip

We show that the data are best fit with a model in which the binary
orbit and circumprimary disk are co-planar. This is consistent with
the suggestion that these systems formed via the monolithic collapse
of a core and subsequent disk fragmentation, which is how massive
binary systems are thought to form
\citep[see][]{Krumholz2009}. However, as of yet, a random association
of disk and binary planes can only be excluded at a 2$\sigma$ level. A
sample of approximately 50 objects is required to reject the random
hypothesis and thus to differentiate between the two scenarios at
3$\sigma$ or higher.

\subsection{Final remarks}

To summarise, we find that spectropolarimetric signatures of young
stellar objects do trace the orientation of their circumstellar disks.
This is independent of a specific mechanism for the linear
polarisation. In itself, this finding is insensitive to the polarising
mechanism as all mechanisms require some form of asymmetric
geometry. We note that scattering in a disk appears a plausible
polarisation mechanism as it naturally explains the relationship
between disk and polarisation angle. Furthermore, assuming the
polarisation of T Tauri stars is due to multiple scattering in
optically thick disks, it is consistent with the observation that T
Tauri star polarisation is generally parallel to imaged disks while
the reverse is true for {\color{black}{{most}}} HAe/Be stars. It is
not clear how optical pumping and polarisation via absorption can
reproduce the different polarimetric behaviour of HAe/Be and T Tauri
stars. Further modelling is required to investigate this issue.

\smallskip

We conclude that our results are entirely consistent with the disks
and orbits of HAe/Be binaries being co-planar, and thus with the scenario
of binary formation via disk fragmentation. Further
spectropolarimetric observations, e.g. provided by SALT, in
conjunction with additional high resolution data, e.g. provided by NIR
interferometry, are required to increase the sample and conclusively
differentiate between aligned and random orientations of disks and
binaries.


\bibliographystyle{aa}
\bibliography{16996}

\begin{thebibliography}{61}
\expandafter\ifx\csname natexlab\endcsname\relax\def\natexlab#1{#1}\fi

\bibitem[{{Akeson} {et~al.}(2003){Akeson}, {Ciardi}, \& {van
  Belle}}]{Akeson2003}
{Akeson}, R.~L., {Ciardi}, D., \& {van Belle}, G.~T. 2003, in SPIE Conf. Ser.,
  ed. {W.~A.~Traub}, Vol. 4838, 1037--1042

\bibitem[{{Akeson} {et~al.}(2002){Akeson}, {Ciardi}, {van Belle}, \&
  {Creech-Eakman}}]{Akeson2002}
{Akeson}, R.~L., {Ciardi}, D.~R., {van Belle}, G.~T., \& {Creech-Eakman}, M.~J.
  2002, ApJ, 566, 1124

\bibitem[{{Baines} {et~al.}(2004){Baines}, {Oudmaijer}, {Mora}, {Eiroa},
  {Porter}, {Mer{\'{\i}}n}, {Montesinos}, {de Winter}, {Collier Cameron},
  {Davies}, {Deeg}, {Ferlet}, {Grady}, {Harris}, {Hoare}, {Horne}, {Lumsden},
  {Miranda}, {Penny}, \& {Quirrenbach}}]{DB2004}
{Baines}, D., {Oudmaijer}, R.~D., {Mora}, A., {et~al.} 2004, MNRAS, 353, 697

\bibitem[{{Baines} {et~al.}(2006){Baines}, {Oudmaijer}, {Porter}, \&
  {Pozzo}}]{DB2006}
{Baines}, D., {Oudmaijer}, R.~D., {Porter}, J.~M., \& {Pozzo}, M. 2006, MNRAS,
  367, 737

\bibitem[{{Bastien} \& {Menard}(1990)}]{Bastien1990}
{Bastien}, P. \& {Menard}, F. 1990, ApJ, 364, 232

\bibitem[{{Bate} {et~al.}(2000){Bate}, {Bonnell}, {Clarke}, {Lubow}, {Ogilvie},
  {Pringle}, \& {Tout}}]{Bate2000}
{Bate}, M.~R., {Bonnell}, I.~A., {Clarke}, C.~J., {et~al.} 2000, MNRAS, 317,
  773

\bibitem[{{Beskrovnaya} {et~al.}(1999){Beskrovnaya}, {Pogodin},
  {Miroshnichenko}, {Th{\'e}}, {Savanov}, {Shakhovskoy}, {Rostopchina},
  {Kozlova}, \& {Kuratov}}]{Besk1999}
{Beskrovnaya}, N.~G., {Pogodin}, M.~A., {Miroshnichenko}, A.~S., {et~al.} 1999,
  A\&A, 343, 163

\bibitem[{{Bonnell} \& {Bate}(2005)}]{Bonnell2005}
{Bonnell}, I.~A. \& {Bate}, M.~R. 2005, MNRAS, 362, 915

\bibitem[{{Cidale} {et~al.}(2001){Cidale}, {Zorec}, \&
  {Tringaniello}}]{Cidale2001}
{Cidale}, L., {Zorec}, J., \& {Tringaniello}, L. 2001, A\&A, 368, 160

\bibitem[{{Clarke} \& {McLean}(1974)}]{Clarke1974}
{Clarke}, D. \& {McLean}, I.~S. 1974, MNRAS, 167, 27P

\bibitem[{{Corder} {et~al.}(2005){Corder}, {Eisner}, \& {Sargent}}]{Corder2005}
{Corder}, S., {Eisner}, J., \& {Sargent}, A. 2005, ApJL, 622, L133

\bibitem[{{Doucet} {et~al.}(2006){Doucet}, {Pantin}, {Lagage}, \&
  {Dullemond}}]{Doucet2006}
{Doucet}, C., {Pantin}, E., {Lagage}, P.~O., \& {Dullemond}, C.~P. 2006, A\&A,
  460, 117

\bibitem[{{Eisner} {et~al.}(2004){Eisner}, {Lane}, {Hillenbrand}, {Akeson}, \&
  {Sargent}}]{Eisner2004}
{Eisner}, J.~A., {Lane}, B.~F., {Hillenbrand}, L.~A., {Akeson}, R.~L., \&
  {Sargent}, A.~I. 2004, ApJ, 613, 1049

\bibitem[{{Fedele} {et~al.}(2008){Fedele}, {van den Ancker}, {Acke}, {van der
  Plas}, {van Boekel}, {Wittkowski}, {Henning}, {Bouwman}, {Meeus}, \&
  {Rafanelli}}]{Fedele2008}
{Fedele}, D., {van den Ancker}, M.~E., {Acke}, B., {et~al.} 2008, A\&A, 491,
  809

\bibitem[{{Fuente} {et~al.}(2006){Fuente}, {Alonso-Albi}, {Bachiller}, {Natta},
  {Testi}, {Neri}, \& {Planesas}}]{Fuente2006}
{Fuente}, A., {Alonso-Albi}, T., {Bachiller}, R., {et~al.} 2006, ApJ, 649, L119

\bibitem[{{Goodwin} {et~al.}(2007){Goodwin}, {Kroupa}, {Goodman}, \&
  {Burkert}}]{Goodwin2007}
{Goodwin}, S.~P., {Kroupa}, P., {Goodman}, A., \& {Burkert}, A. 2007, in
  Protostars and Planets V, ed. B.~{Reipurth}, D.~{Jewitt}, \& K.~{Keil},
  133--147

\bibitem[{{Harrington} \& {Kuhn}(2007)}]{Harrington2007}
{Harrington}, D.~M. \& {Kuhn}, J.~R. 2007, ApJL, 667, L89

\bibitem[{{Harrington} \& {Kuhn}(2009)}]{Harrington2009a}
{Harrington}, D.~M. \& {Kuhn}, J.~R. 2009, ApJS, 180, 138

\bibitem[{{Herbig}(1960)}]{Herbig1960}
{Herbig}, G.~H. 1960, ApJ Supplement, 4, 337

\bibitem[{{Hern\'{a}ndez} {et~al.}(2004){Hern\'{a}ndez}, {Calvet},
  {Brice\~{n}o}, {Hartmann}, \& {Berlind}}]{Hernandez2004}
{Hern\'{a}ndez}, J., {Calvet}, N., {Brice\~{n}o}, C., {Hartmann}, L., \&
  {Berlind}, P. 2004, AJ, 127, 1682

\bibitem[{{Hillenbrand} {et~al.}(1992){Hillenbrand}, {Strom}, {Vrba}, \&
  {Keene}}]{Hillenbrand1992}
{Hillenbrand}, L.~A., {Strom}, S.~E., {Vrba}, F.~J., \& {Keene}, J. 1992,
  Astrophysical Journal, 397, 613

\bibitem[{{Jensen} {et~al.}(2004){Jensen}, {Mathieu}, {Donar}, \&
  {Dullighan}}]{Jensen2004}
{Jensen}, E.~L.~N., {Mathieu}, R.~D., {Donar}, A.~X., \& {Dullighan}, A. 2004,
  ApJ, 600, 789

\bibitem[{{Kitamura} {et~al.}(2002){Kitamura}, {Momose}, {Yokogawa}, {Kawabe},
  {Tamura}, \& {Ida}}]{Kitamura2002}
{Kitamura}, Y., {Momose}, M., {Yokogawa}, S., {et~al.} 2002, ApJ, 581, 357

\bibitem[{{Kraus} {et~al.}(2008){Kraus}, {Preibisch}, \& {Ohnaka}}]{Kraus2008}
{Kraus}, S., {Preibisch}, T., \& {Ohnaka}, K. 2008, ApJ, 676, 490

\bibitem[{{Krumholz} {et~al.}(2009){Krumholz}, {Klein}, {McKee}, {Offner}, \&
  {Cunningham}}]{Krumholz2009}
{Krumholz}, M.~R., {Klein}, R.~I., {McKee}, C.~F., {Offner}, S.~S.~R., \&
  {Cunningham}, A.~J. 2009, Sci, 323, 754

\bibitem[{{Kuhn} {et~al.}(2007){Kuhn}, {Berdyugina}, {Fluri}, {Harrington}, \&
  {Stenflo}}]{Kuhn2007}
{Kuhn}, J.~R., {Berdyugina}, S.~V., {Fluri}, D.~M., {Harrington}, D.~M., \&
  {Stenflo}, J.~O. 2007, ApJL, 668, L63

\bibitem[{{Kuhn} {et~al.}(2010){Kuhn}, {Geiss}, \& {Harrington}}]{Kuhn2010}
{Kuhn}, J.~R., {Geiss}, B., \& {Harrington}, D.~M. 2010, ArXiv e-prints

\bibitem[{{Maheswar} {et~al.}(2002){Maheswar}, {Manoj}, \&
  {Bhatt}}]{GMaheswar2002}
{Maheswar}, G., {Manoj}, P., \& {Bhatt}, H.~C. 2002, A\&A, 387, 1003

\bibitem[{{Malbet} {et~al.}(2005){Malbet}, {Lachaume}, {Berger}, {Colavita},
  {di Folco}, {Eisner}, {Lane}, {Millan-Gabet}, {S{\'e}gransan}, \&
  {Traub}}]{Malbet2005}
{Malbet}, F., {Lachaume}, R., {Berger}, J., {et~al.} 2005, A\&A, 437, 627

\bibitem[{{Mannings} \& {Sargent}(1997)}]{Mannings1997}
{Mannings}, V. \& {Sargent}, A.~I. 1997, ApJ, 490, 792

\bibitem[{{Manoj} {et~al.}(2007){Manoj}, {Ho}, {Ohashi}, {Zhang}, {Hasegawa},
  {Chen}, {Bhatt}, \& {Ashok}}]{Manoj2007}
{Manoj}, P., {Ho}, P.~T.~P., {Ohashi}, N., {et~al.} 2007, ApJL, 667, L187

\bibitem[{{Mathieu} {et~al.}(1995){Mathieu}, {Adams}, {Fuller}, {Jensen},
  {Koerner}, \& {Sargent}}]{Mathieu1995}
{Mathieu}, R.~D., {Adams}, F.~C., {Fuller}, G.~A., {et~al.} 1995, AJ, 109, 2655

\bibitem[{{McLean} \& {Clarke}(1979)}]{McLean1979}
{McLean}, I.~S. \& {Clarke}, D. 1979, MNRAS, 186, 245

\bibitem[{{Moeckel} \& {Bally}(2007)}]{JBally2007}
{Moeckel}, N. \& {Bally}, J. 2007, ApJ, 656, 275

\bibitem[{{Monin} {et~al.}(2006){Monin}, {M\'{e}nard}, \&
  {Peretto}}]{Monin2006}
{Monin}, J.-L., {M\'{e}nard}, F., \& {Peretto}, N. 2006, A\&A, 446, 201

\bibitem[{{Monnier} {et~al.}(2006){Monnier}, {Berger}, {Millan-Gabet}, {Traub},
  {Schloerb}, {Pedretti}, {Benisty}, {Carleton}, {Haguenauer}, {Kern},
  {Labeye}, {Lacasse}, {Malbet}, {Perraut}, {Pearlman}, \&
  {Zhao}}]{Monnier2006}
{Monnier}, J.~D., {Berger}, J.-P., {Millan-Gabet}, R., {et~al.} 2006, ApJ, 647,
  444

\bibitem[{{Monnier} {et~al.}(2008){Monnier}, {Tannirkulam}, {Tuthill},
  {Ireland}, {Cohen}, {Danchi}, \& {Baron}}]{Monnier2008}
{Monnier}, J.~D., {Tannirkulam}, A., {Tuthill}, P.~G., {et~al.} 2008, ApJL,
  681, L97

\bibitem[{{Monnier}(2005)}]{Monnier2005}
{Monnier}, J.~D. e.~a. 2005, ApJ, 624, 832

\bibitem[{{Mottram} {et~al.}(2007){Mottram}, {Vink}, {Oudmaijer}, \&
  {Patel}}]{JCMottram2007}
{Mottram}, J.~C., {Vink}, J.~S., {Oudmaijer}, R.~D., \& {Patel}, M. 2007,
  MNRAS, 377, 1363

\bibitem[{{Natta} {et~al.}(2001){Natta}, {Prusti}, {Neri}, {Wooden}, {Grinin},
  \& {Mannings}}]{Natta2001}
{Natta}, A., {Prusti}, T., {Neri}, R., {et~al.} 2001, A\&A, 371, 186

\bibitem[{{Okamoto} {et~al.}(2009){Okamoto}, {Kataza}, {Honda}, {Fujiwara},
  {Momose}, {Ohashi}, {Fujiyoshi}, {Sakon}, {Sako}, {Yamashita}, {Miyata}, \&
  {Onaka}}]{Okamoto2009}
{Okamoto}, Y.~K., {Kataza}, H., {Honda}, M., {et~al.} 2009, ApJ, 706, 665

\bibitem[{{Oudmaijer} \& {Drew}(1999)}]{Oudmaijer1999}
{Oudmaijer}, R.~D. \& {Drew}, J.~E. 1999, MNRAS, 305, 166

\bibitem[{{Patel} {et~al.}(2006){Patel}, {Oudmaijer}, {Vink}, {Mottram}, \&
  {Davies}}]{Patel2006}
{Patel}, M., {Oudmaijer}, R.~D., {Vink}, J.~S., {Mottram}, J.~C., \& {Davies},
  B. 2006, MNRAS, 373, 1641

\bibitem[{{Pirzkal} {et~al.}(1997){Pirzkal}, {Spillar}, \&
  {Dyck}}]{Pirzkal1997}
{Pirzkal}, N., {Spillar}, E.~J., \& {Dyck}, H.~M. 1997, ApJ, 481, 392

\bibitem[{{Poeckert}(1975)}]{Poeckert1975}
{Poeckert}, R. 1975, ApJ, 196, 777

\bibitem[{{Poeckert} \& {Marlborough}(1976)}]{Poeckert1976}
{Poeckert}, R. \& {Marlborough}, J.~M. 1976, ApJ, 206, 182

\bibitem[{{Preibisch} {et~al.}(2006){Preibisch}, {Kraus}, {Driebe}, {van
  Boekel}, \& {Weigelt}}]{Preibisch2006}
{Preibisch}, T., {Kraus}, S., {Driebe}, T., {van Boekel}, R., \& {Weigelt}, G.
  2006, A\&A, 458, 235

\bibitem[{{Rousselet-Perraut} {et~al.}(2010){Rousselet-Perraut}, {Benisty},
  {Mourard}, {Rajabi}, {Bacciotti}, {B{\'e}rio}, {Bonneau}, {Chesneau},
  {Clausse}, {Delaa}, {Marcotto}, {Roussel}, {Spang}, {Stee}, {Tallon-Bosc},
  {McAlister}, {Ten Brummelaar}, {Sturmann}, {Sturmann}, {Turner},
  {Farrington}, \& {Goldfinger}}]{Perraut2010}
{Rousselet-Perraut}, K., {Benisty}, M., {Mourard}, D., {et~al.} 2010, A\&A,
  516, L1

\bibitem[{{Stecklum} {et~al.}(1995){Stecklum}, {Eckart}, {Henning}, \&
  {Loewe}}]{Stecklum1995}
{Stecklum}, B., {Eckart}, A., {Henning}, T., \& {Loewe}, M. 1995, A\&A, 296,
  463

\bibitem[{{Th\'{e}} {et~al.}(1994){Th\'{e}}, {de Winter}, \& {Perez}}]{The1994}
{Th\'{e}}, P.~S., {de Winter}, D., \& {Perez}, M.~R. 1994, A\&AS, 104, 315

\bibitem[{{Thomas} {et~al.}(2007){Thomas}, {van der Bliek}, {Rodgers},
  {Doppmann}, \& {Bouvier}}]{Thomas2007}
{Thomas}, S.~J., {van der Bliek}, N.~S., {Rodgers}, B., {Doppmann}, G., \&
  {Bouvier}, J. 2007, in IAU Symp, Vol. 240, IAU Symp., ed. W.~I. {Hartkopf},
  E.~F. {Guinan}, \& P.~{Harmanec}, 250--253

\bibitem[{{Tody}(1993)}]{IRAF}
{Tody}, D. 1993, in A.S.P. Conf.Ser., Vol.~52, Astronomical Data Analysis
  Software and Systems II, ed. R.~J. {Hanisch}, R.~J.~V. {Brissenden}, \&
  J.~{Barnes}, 173--183

\bibitem[{{Vink} {et~al.}(2002){Vink}, {Drew}, {Harries}, \&
  {Oudmaijer}}]{Vink2002}
{Vink}, J.~S., {Drew}, J.~E., {Harries}, T.~J., \& {Oudmaijer}, R.~D. 2002,
  MNRAS, 337, 356

\bibitem[{{Vink} {et~al.}(2005{\natexlab{a}}){Vink}, {Drew}, {Harries},
  {Oudmaijer}, \& {Unruh}}]{Vink2005a}
{Vink}, J.~S., {Drew}, J.~E., {Harries}, T.~J., {Oudmaijer}, R.~D., \& {Unruh},
  Y. 2005{\natexlab{a}}, MNRAS, 359, 1049

\bibitem[{{Vink} {et~al.}(2005{\natexlab{b}}){Vink}, {Harries}, \&
  {Drew}}]{DisksVink2005}
{Vink}, J.~S., {Harries}, T.~J., \& {Drew}, J.~E. 2005{\natexlab{b}}, A\&A,
  430, 213

\bibitem[{{Vink} {et~al.}(2005{\natexlab{c}}){Vink}, {O'Neill}, {Els}, \&
  {Drew}}]{Vink2005b}
{Vink}, J.~S., {O'Neill}, P.~M., {Els}, S.~G., \& {Drew}, J.~E.
  2005{\natexlab{c}}, A\&A, 438, L21

\bibitem[{{Waters} \& {Waelkens}(1998)}]{Waters1998}
{Waters}, L.~B.~F.~M. \& {Waelkens}, C. 1998, ARA\&A, 36, 233

\bibitem[{{Weigelt} {et~al.}(2002){Weigelt}, {Balega}, {Hofmann}, \&
  {Preibisch}}]{Weigelt2002}
{Weigelt}, G., {Balega}, Y.~Y., {Hofmann}, K.-H., \& {Preibisch}, T. 2002,
  A\&A, 392, 937

\bibitem[{{Wheelwright} {et~al.}(2010){Wheelwright}, {Oudmaijer}, \&
  {Goodwin}}]{Wheelwright2010}
{Wheelwright}, H.~E., {Oudmaijer}, R.~D., \& {Goodwin}, S.~P. 2010, MNRAS, 401,
  1199

\bibitem[{{Wolf} {et~al.}(2001){Wolf}, {Stecklum}, \& {Henning}}]{SWolf2001}
{Wolf}, S., {Stecklum}, B., \& {Henning}, T. 2001, in IAU Symp., Vol. 200, The
  Formation of Binary Stars, ed. H.~{Zinnecker} \& R.~{Mathieu}, 295--304

\bibitem[{Zinnecker \& Yorke(2007)}]{ZinneckerandYorke2007}
Zinnecker, H. \& Yorke, H. 2007, ARA\&A, 45, 481

\end{thebibliography}

\appendix

\section{Observed spectropolarimetric signatures}

\label{spec_pol_app}

\begin{center}
  \begin{figure*}
    \begin{center}
      \begin{tabular} {p{0.3\textwidth} p{0.3\textwidth} p{0.3\textwidth}}

	\includegraphics[width=0.3\textwidth]{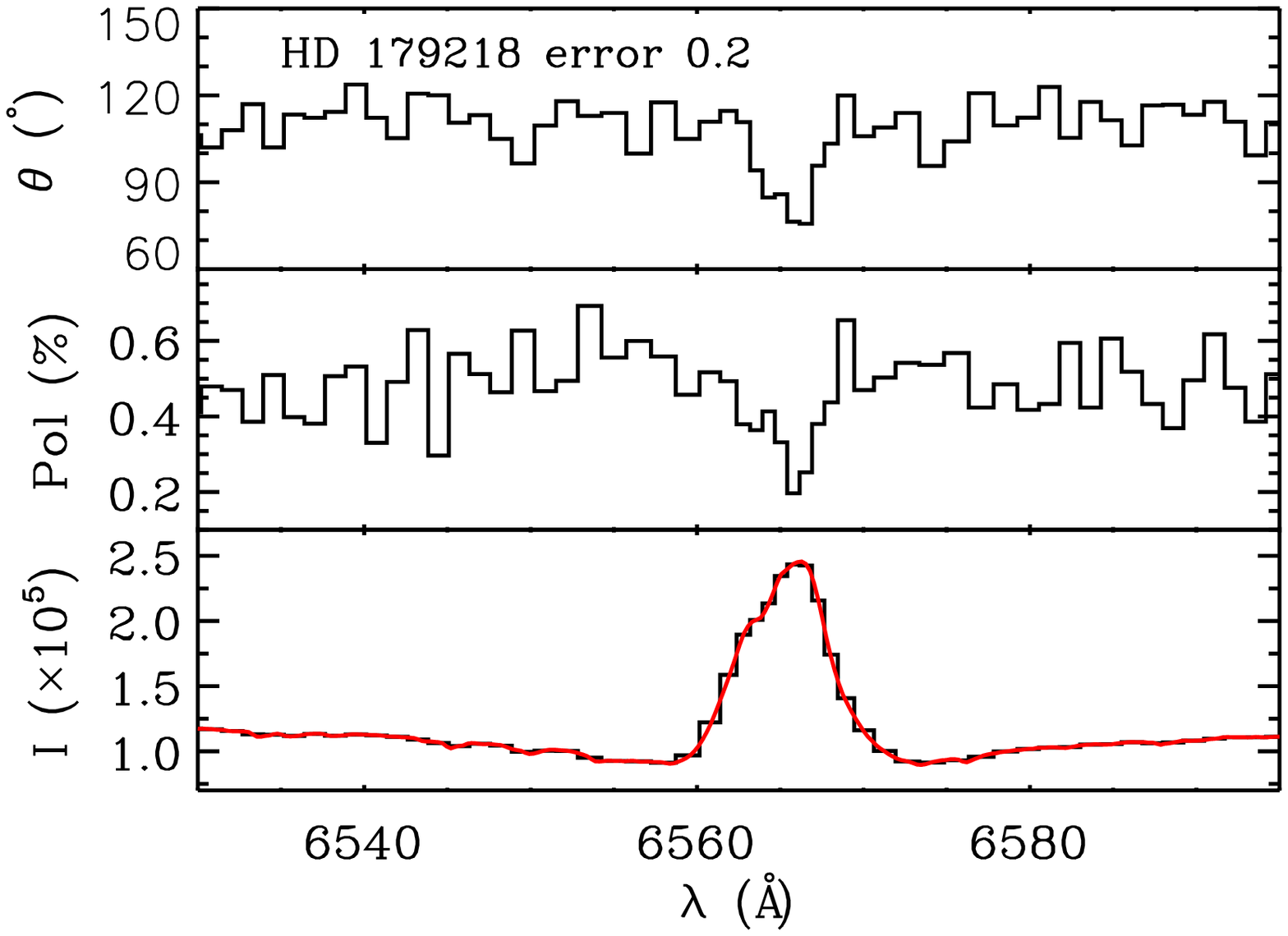} & 
	\includegraphics[width=0.3\textwidth]{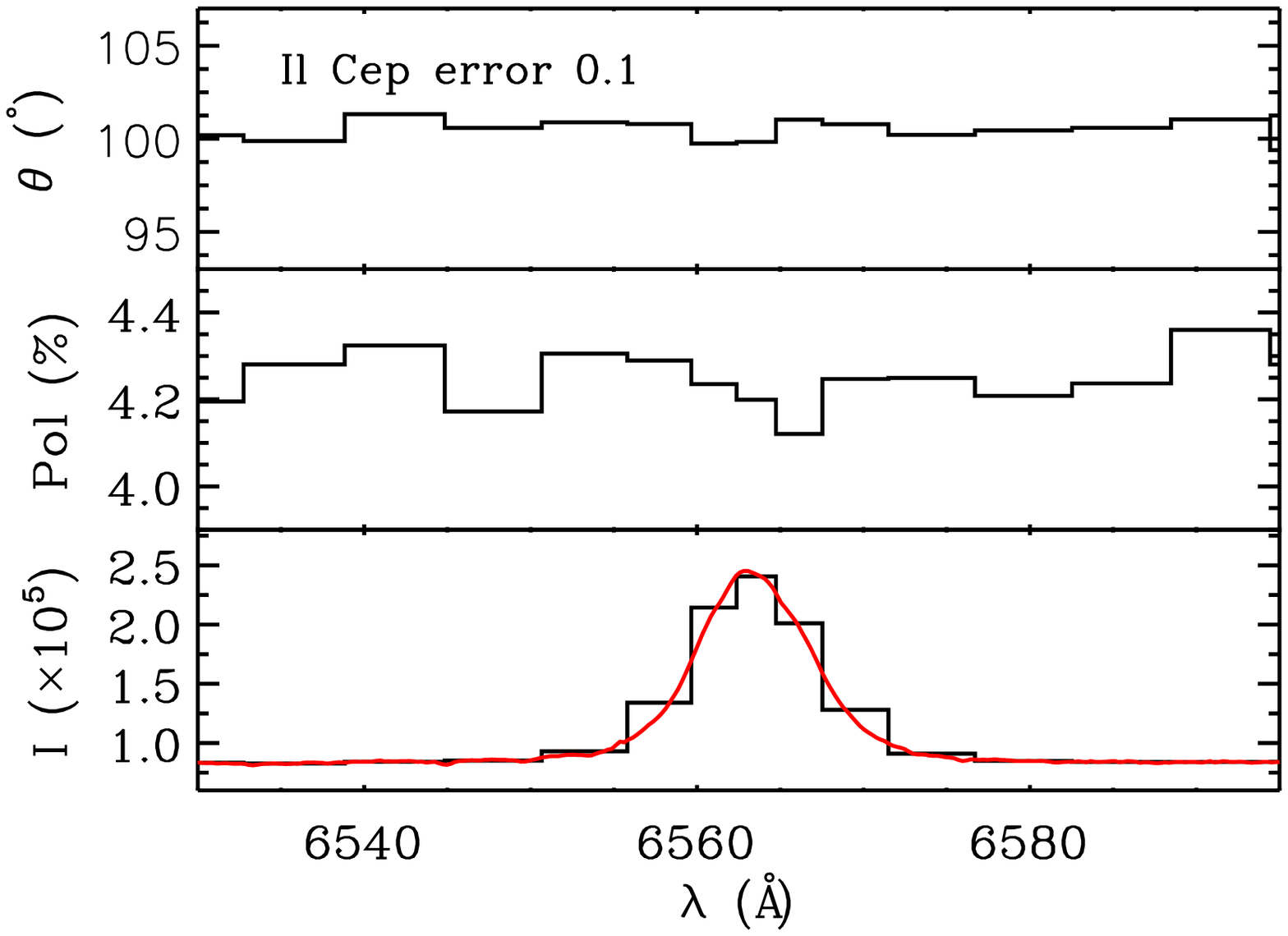} &
	\includegraphics[width=0.3\textwidth]{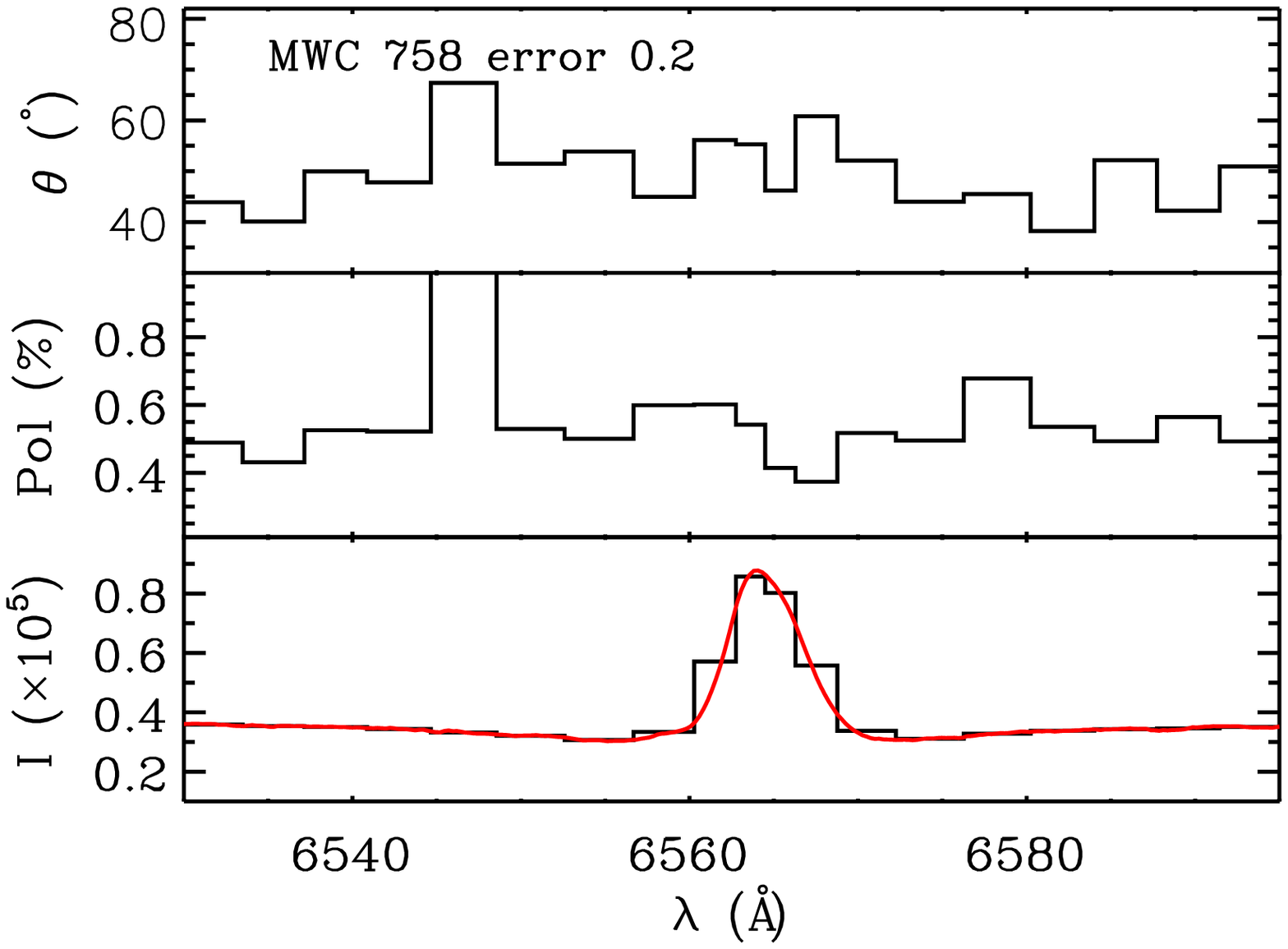} \\ 
	\includegraphics[width=0.3\textwidth]{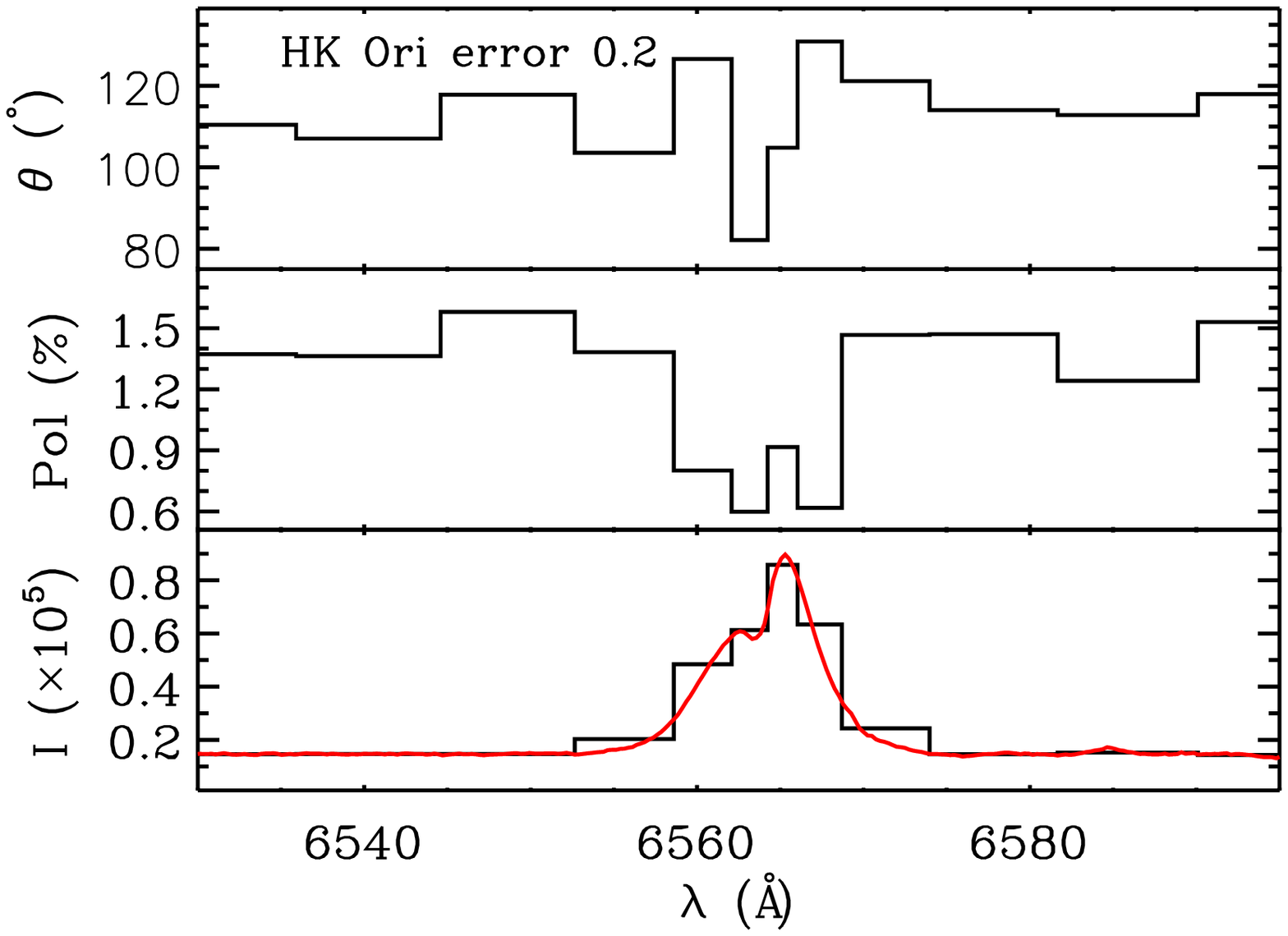} &
	\includegraphics[width=0.3\textwidth]{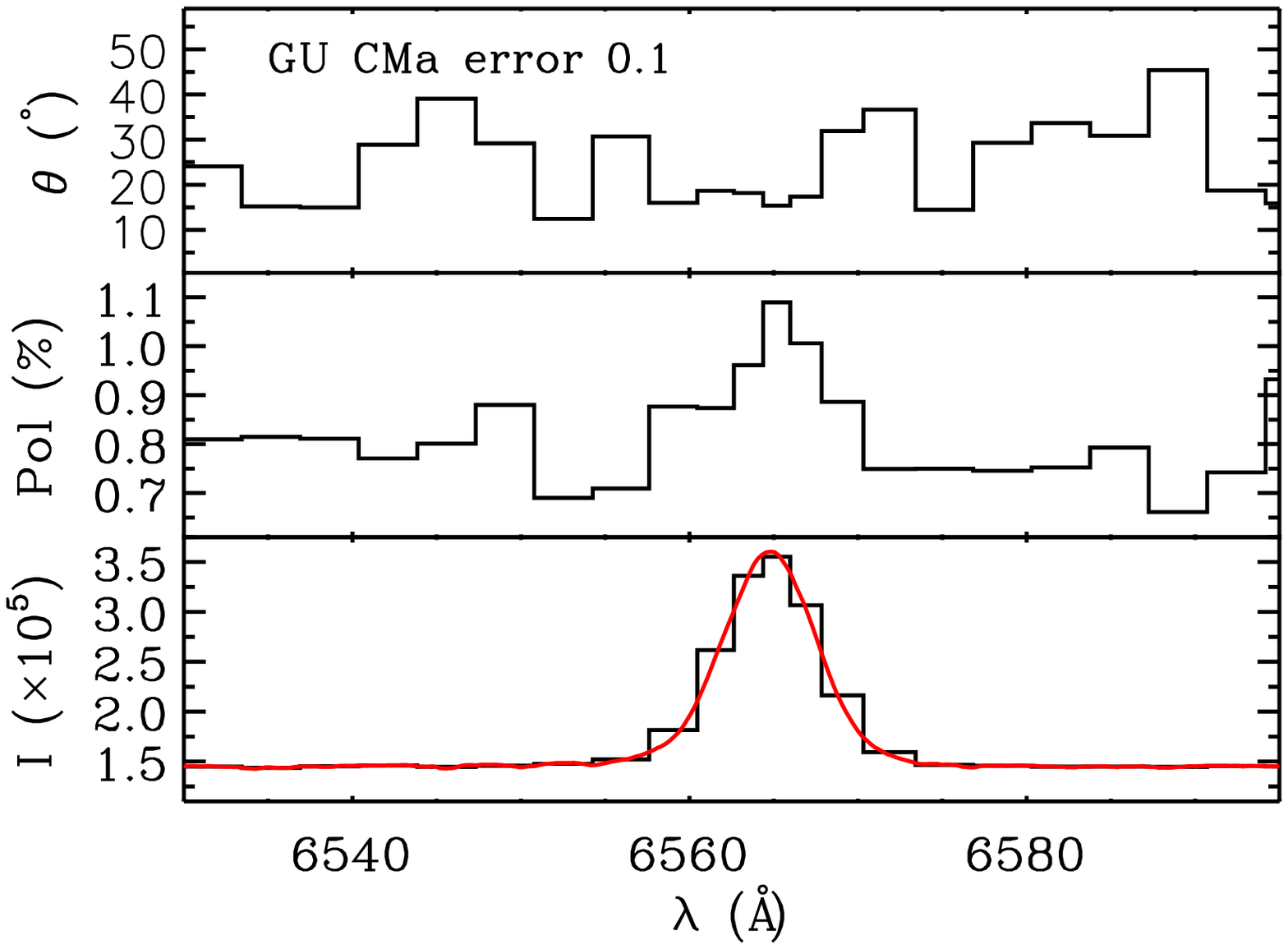} & 
	\includegraphics[width=0.3\textwidth]{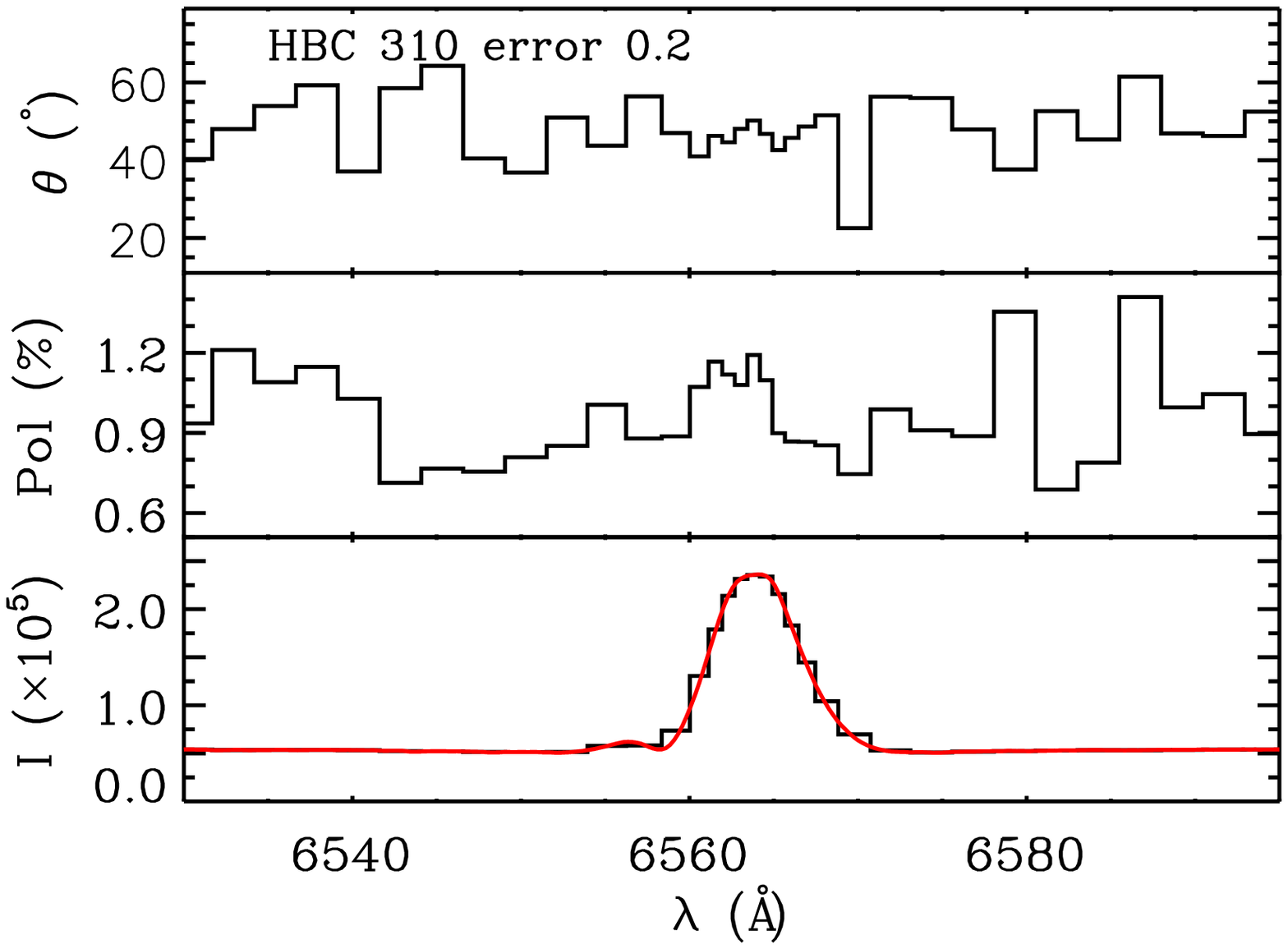} \\
	\includegraphics[width=0.3\textwidth]{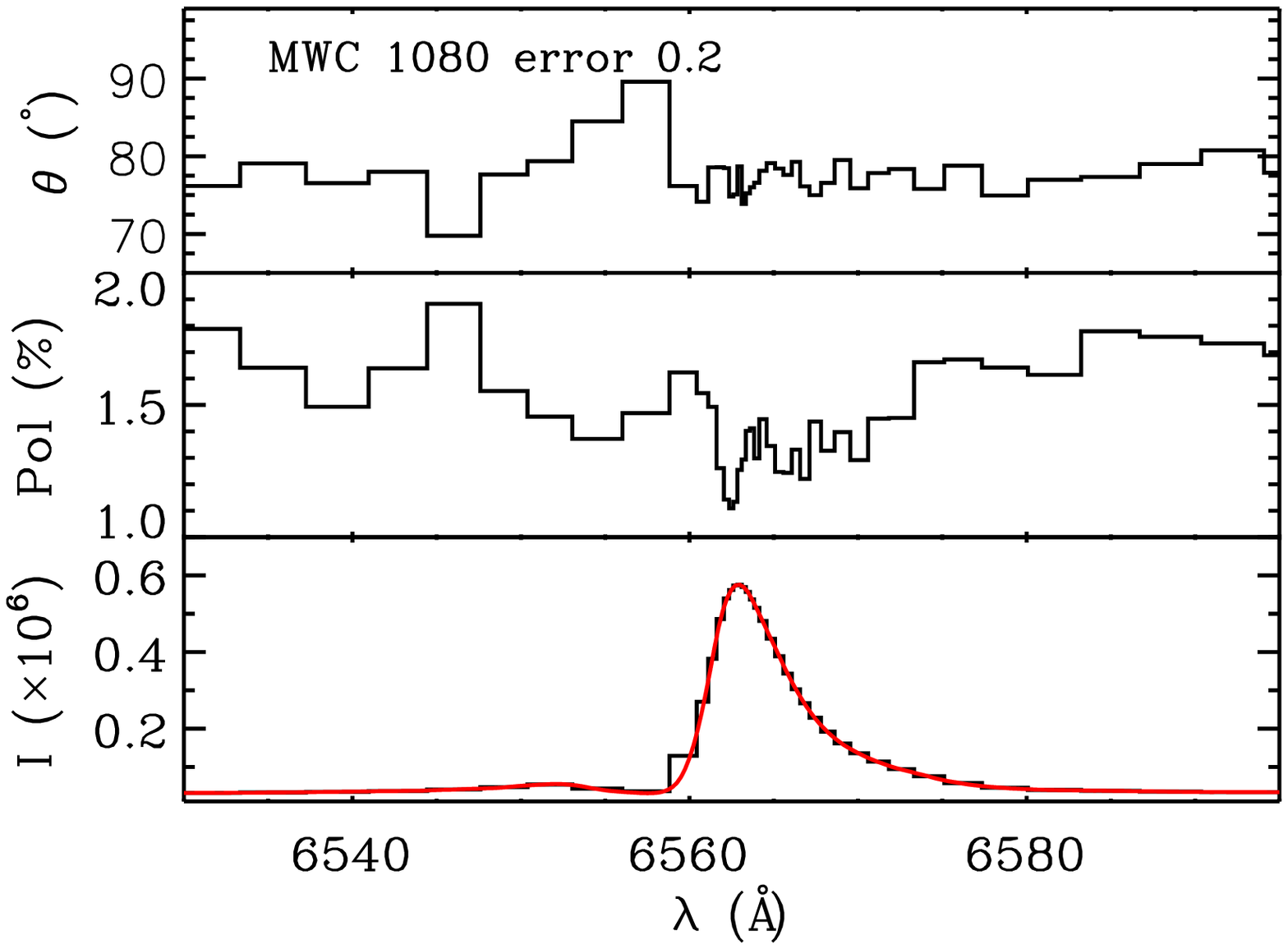} & 
	\includegraphics[width=0.3\textwidth]{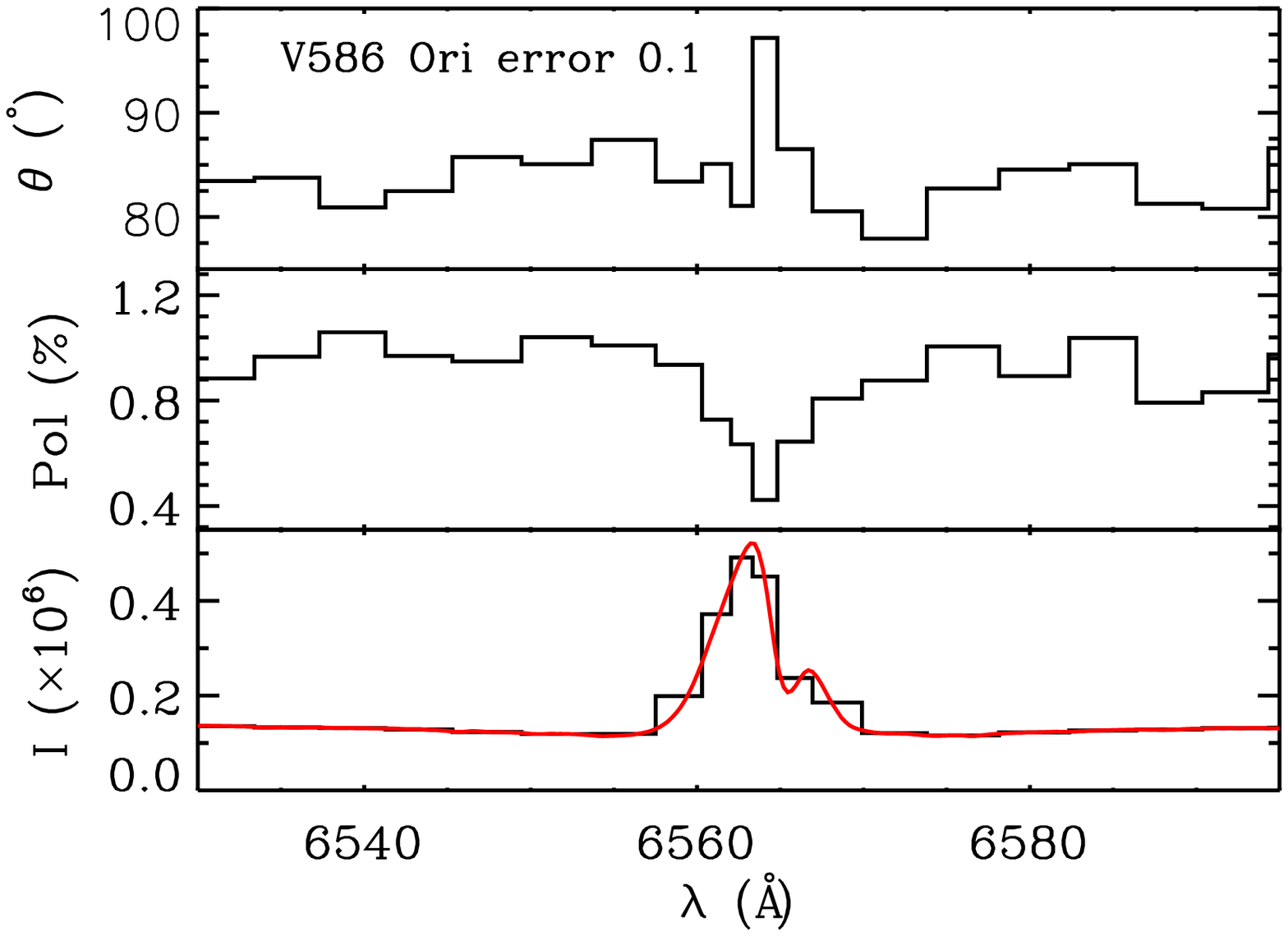} &
	\includegraphics[width=0.3\textwidth]{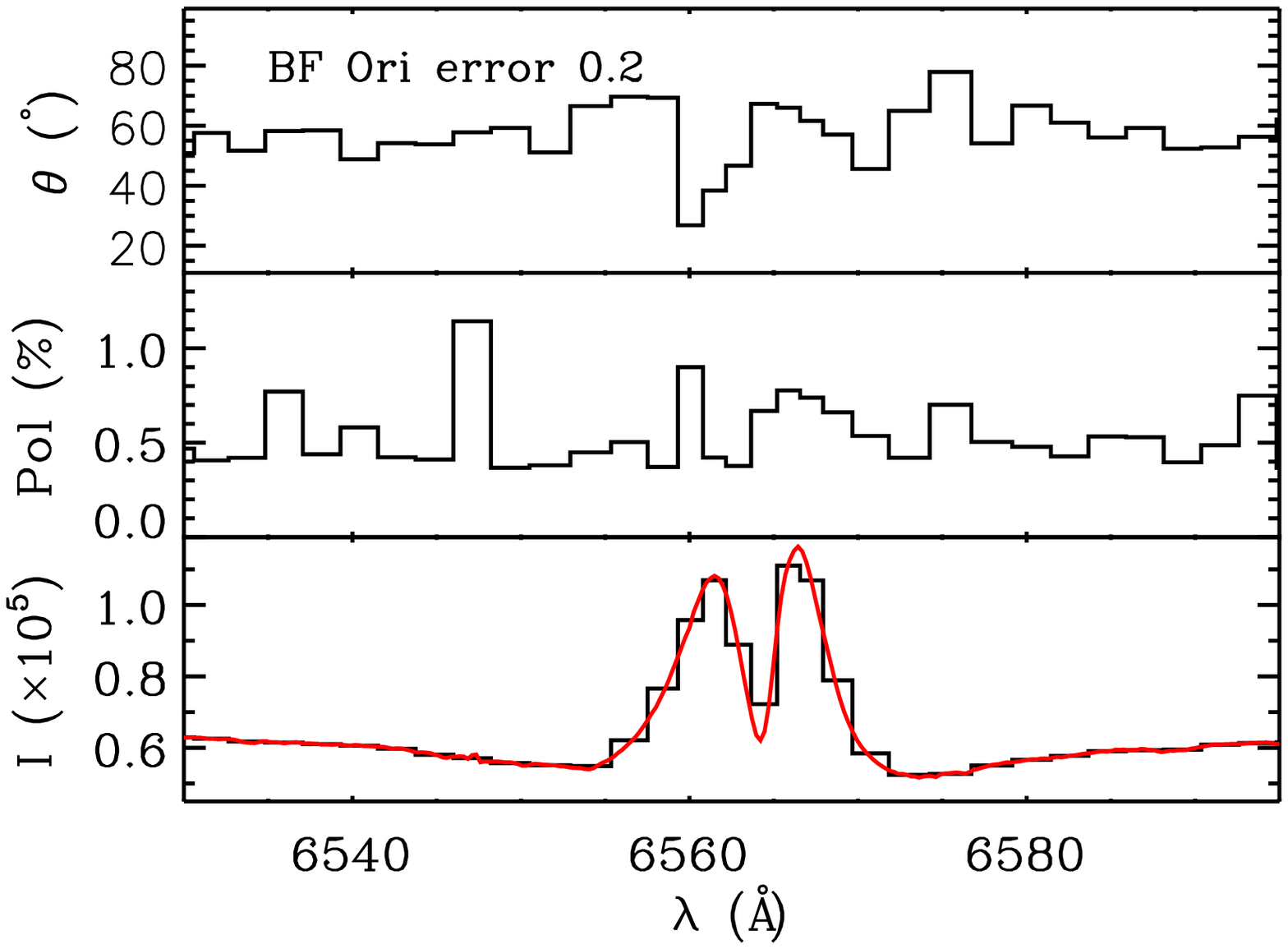} \\ 
	\includegraphics[width=0.3\textwidth]{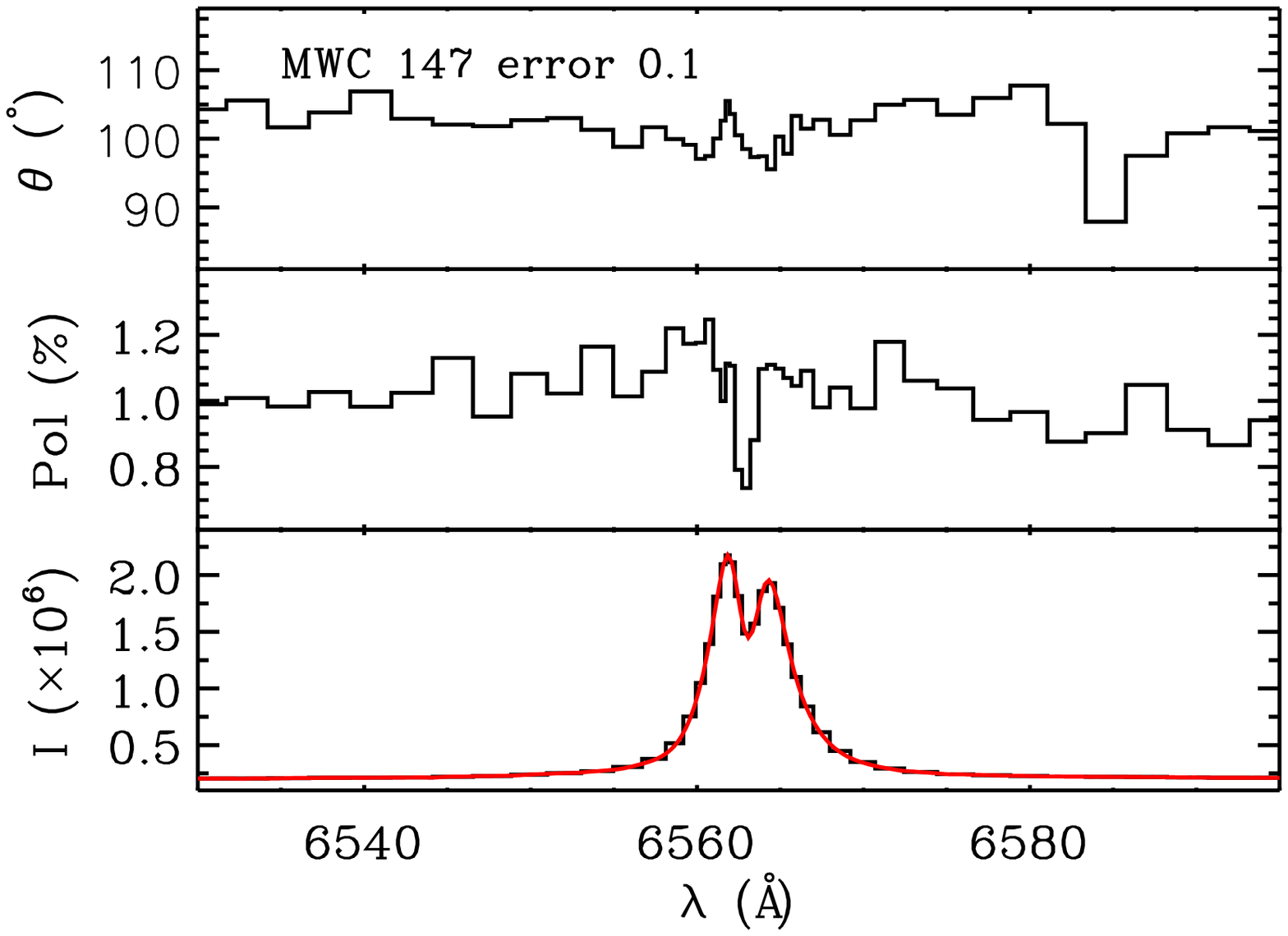} &
	\includegraphics[width=0.3\textwidth]{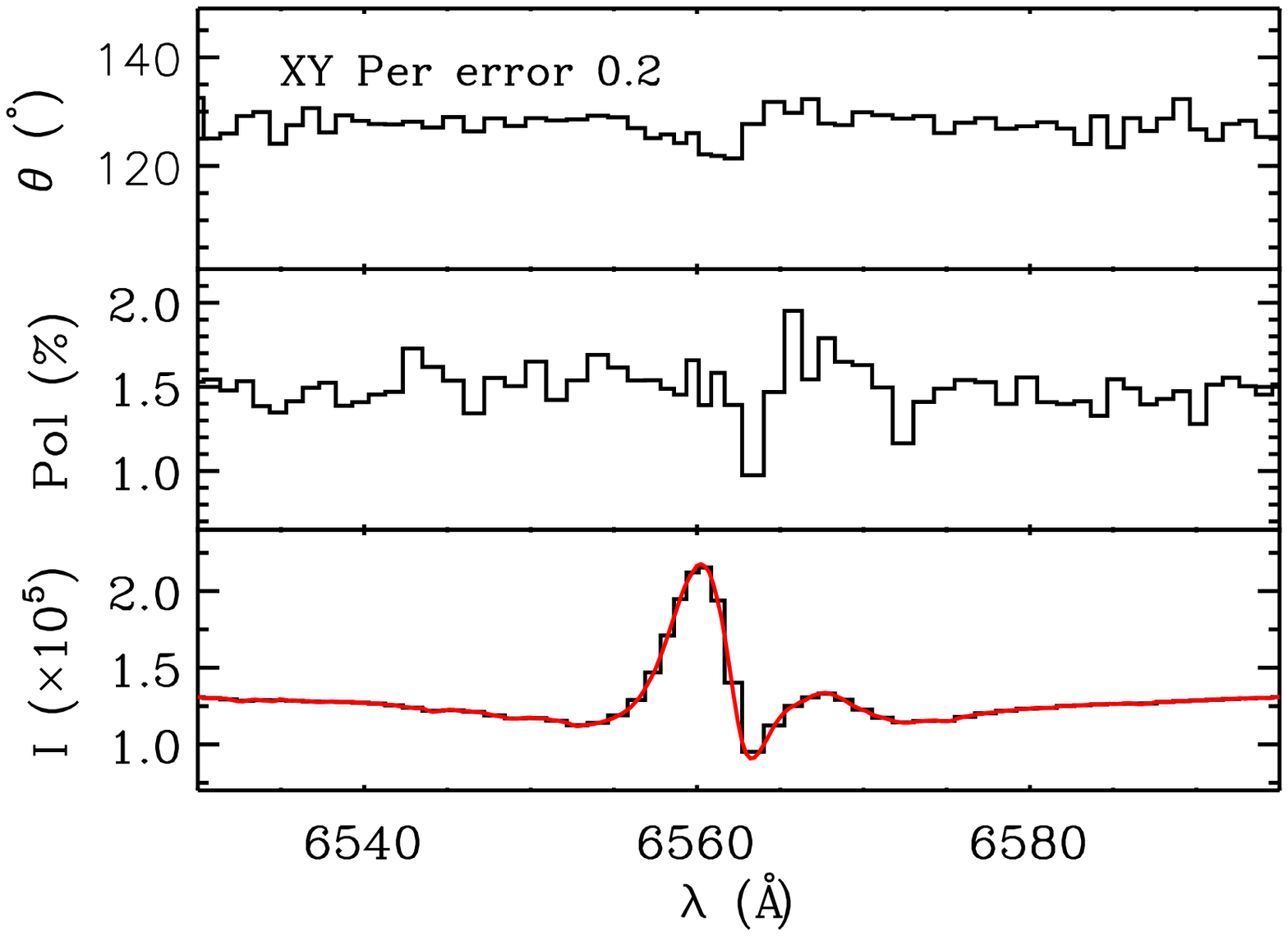} & 
	\includegraphics[width=0.3\textwidth]{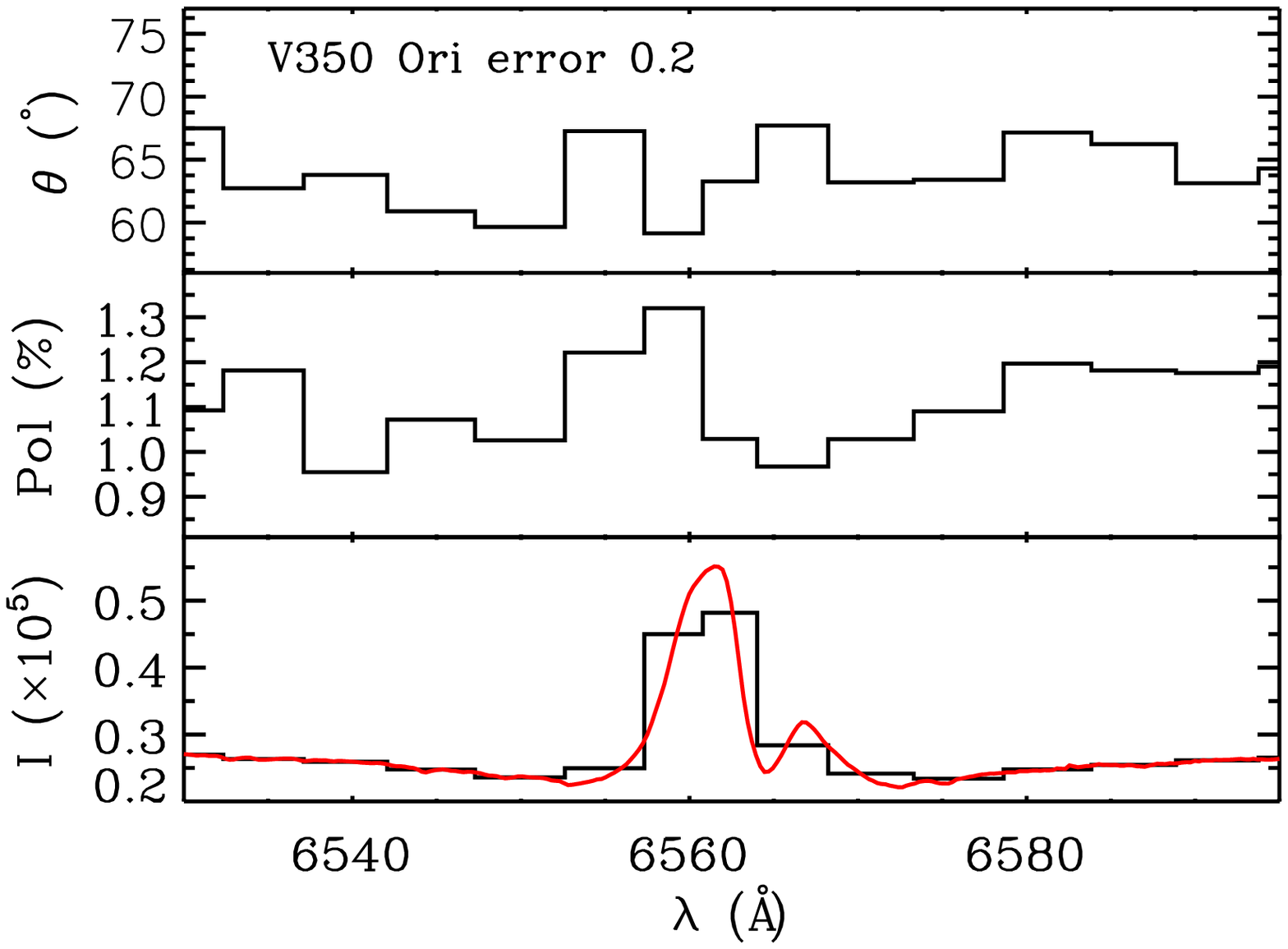} \\

      \end{tabular}
      
      \caption{The spectropolarimetric signatures of the sample. For
  each object the spectropolarimetric PA, the percentage polarisation,
  and the Stokes intensity spectra are presented centred upon
  H$\alpha$. The data are binned to a constant {\color{black}{{polarisation}}} error, which is stated in the plots. The solid red line is the un-binned line
  profile.\label{spec_pol_tri}}

    \end{center}  
  \end{figure*}
\end{center}

\begin{center}
  \begin{figure*}
    \begin{center}
      \begin{tabular} {c c}

	\includegraphics[width=0.3\textwidth]{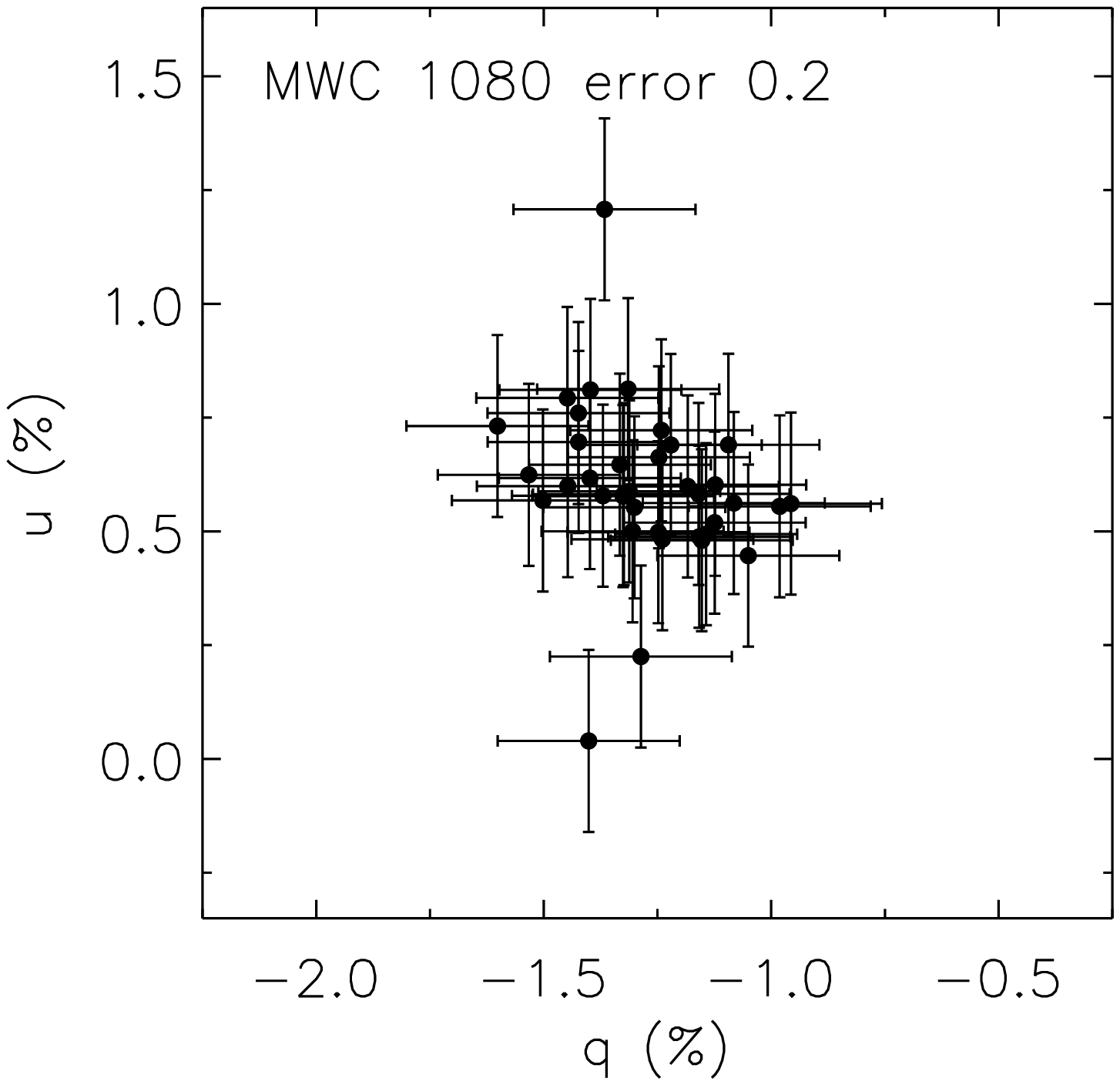} & 
	\includegraphics[width=0.3\textwidth]{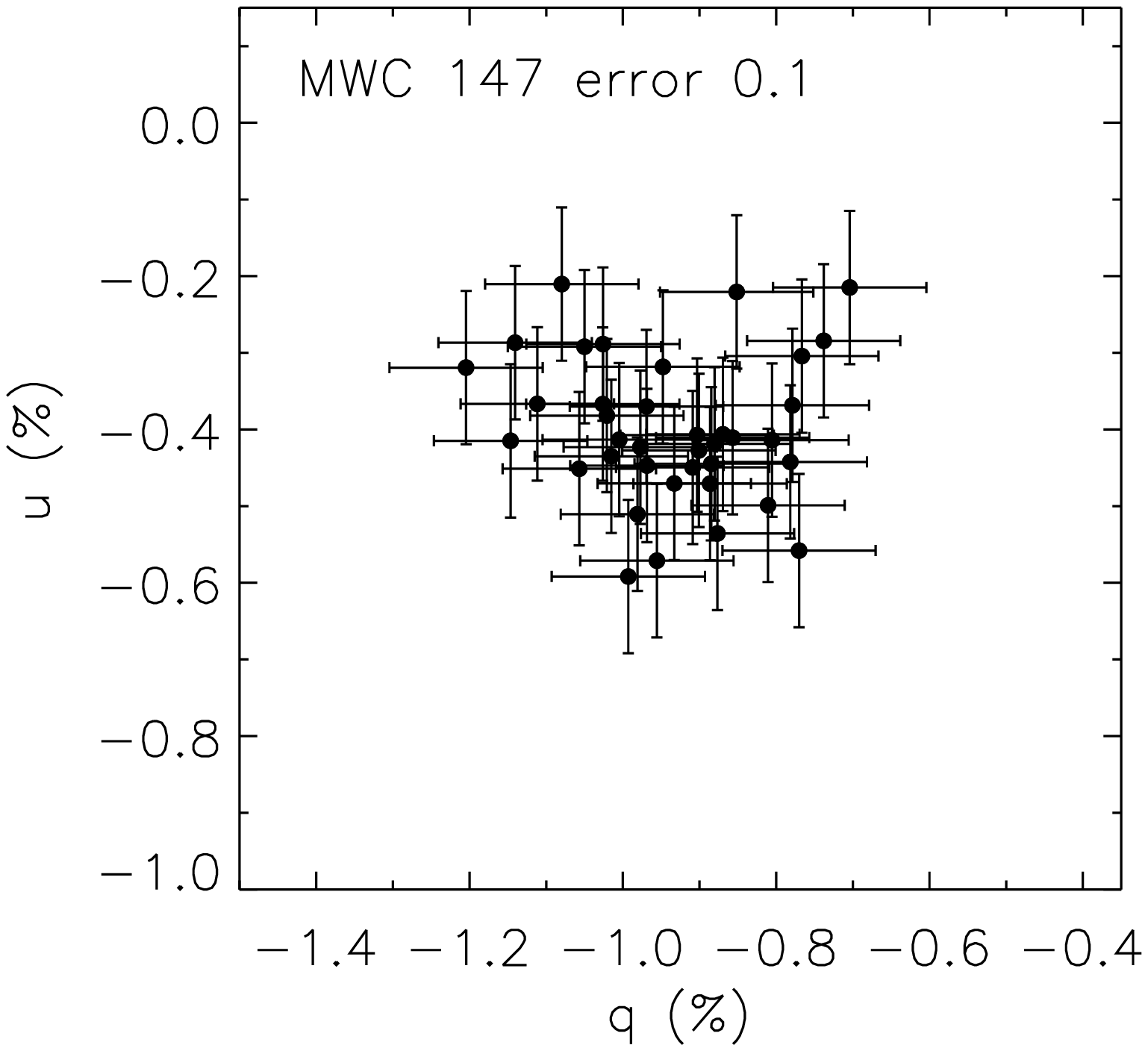} \\

      \end{tabular}
      
      \caption{The $QU$ diagrams of MWC 1080 and MWC 147. The data are binned to a constant {\color{black}{{polarisation}}} error, which is stated in the plots. \label{qud}}

    \end{center}  
  \end{figure*}
\end{center}

Here we present the spectropolarimetric signatures of the entire sample over
H$\alpha$. Figure \ref{spec_pol_tri} presents the signatures in terms
of the total flux, the amount of polarisation and the polarisation
angle over the H$\alpha$ line. {\color{black}{{The $QU$ diagrams of MWC 1080 and MWC 147, objects which display a spectropolarimetric signature that was not presented in the main body of the paper, are presented in Figure \ref{qud}.}}}

\end{document}